\documentclass[aip,rsi,reprint]{revtex4-1}

\usepackage{graphicx}

\usepackage[colorlinks=true,linkcolor=blue]{hyperref}%

\newcommand{\lsmo}{La$_{1.05}$Sr$_{1.95}$Mn$_2$O$_7$}

\begin{document}

\title{RASOR: An advanced instrument for soft x-ray reflectivity and diffraction}%

\author{T.A.W. Beale}
\homepage[Visit: ]{http://www.dur.ac.uk/xray.magnetism/rasor}
\affiliation{Department of Physics, Durham University, South Road, Durham, DH1 3LE, UK} 
\author{T.P.A. Hase}
\affiliation{Department of Physics, University of Warwick, Coventry, CV4 7AL, UK}

\author{T. Iida}
\author{K. Endo}
\affiliation{Toyama Co. Ltd., 4-13-16 Hibarigaoka 4-chome, Zama-shi, Kanagawa Prefecture, Japan 252-0003}

\author{P. Steadman}
\author{A.R. Marshall}
\author{S.S. Dhesi}
\author{G. van der Laan}
\affiliation{Diamond Light Source Ltd., Harwell Science and Innovation Campus, Chilton, Didcot, Oxfordshire, OX11 0DE, UK}

\author{P.D. Hatton}
\email{p.d.hatton@dur.ac.uk}
\affiliation{Department of Physics, Durham University, South Road, Durham, DH1 3LE, UK}%

\date{October 2009}%
\revised{\today}%
\begin{abstract}

We report the design and construction of a novel soft x-ray diffractometer installed at Diamond Light Source.   The beamline endstation RASOR is constructed for general users and designed primarily for the study of single crystal diffraction and thin film reflectivity.   The instrument is comprised of a limited three circle ($\theta$, $2\theta$, $\chi$) diffractometer with an additional removable rotation ($\phi$) stage.    It is equipped with a liquid helium cryostat, and post-scatter polarization analysis.   Motorised motions are provided for the precise positioning of the sample onto the diffractometer centre of rotation, and for positioning the centre of rotation onto the x-ray beam.   The functions of the instrument have been tested at Diamond Light Source, and initial test measurements are provided, demonstrating the potential of the instrument.

\end{abstract}
\maketitle

\section{Introduction}

Resonant x-ray magnetic scattering has played a vital role in the understanding of correlated electron systems for more than thirty years.   Despite the extensive research in this technique, the vast majority of experiments have observed resonances in the hard x-ray regime (5-15~keV).     However in recent years it has become increasingly apparent that direct observations of the magnetic order must be undertaken through using a resonant transition that excites into the magnetically ordered electronic band.   Thus for transition metals, the $L_{2,3}$ edges provide a direct probe of the magnetically active $3d$ electron band\cite{Laan:120, Laan:570}, and the rare earth $M_{4,5}$ edges probe the $4f$ electrons\cite{Spencer:1725}.   Diffraction utilising such transitions, known as soft x-ray resonant diffraction, has provided the community with a wealth of results from a variety of systems, and is now becoming a standard technique for physicists studying correlated electron systems.

In this article, we describe the endstation RASOR, (Reflectivity and Advanced Scattering from Ordered Regimes), that has been constructed to facilitate soft x-ray diffraction and reflectivity experiments at Diamond Light Source.   It has been designed to be used by the scattering community, by providing an extendable endstation that can be used as a standard instrument but also has the capability of supporting future upgrades.

Soft x-ray scattering experiments were first undertaken over twenty years ago, with an in-vacuum diffractometer designed for studying multilayer films\cite{Jark:654}.  The first scattering experiment at a transition metal $L$ edge was performed by Kao \emph{et al.} \cite{kao:373}, on a single crystal of Fe(110), directly testing the prediction by Hannon of a large resonant enhancement\cite{Hannon:1245}.   By measuring the reflectivity from the sample, and constructing a model of the reflectivity, they were able to demonstrate that their measurements were sensitive to the magnetic parameters calculated for iron.  The experiment was proceeded by a study of Co thin films\cite{Kao:9599} with circular polarized light, performed on the AT\&T Bell Laboratories Dragon beamline\cite{Chen:119} at the NSLS. By combining the dichroic effects of circular polarized light with magnetic reflectivity, asymmetry ratios were obtainable for weak magnetic moments, as a function of depth in the material, demonstrating that the technique could be used to separate the magnetic and structural roughness of thin films.   

Tonnerre and coworkers pioneered the use of this technique with layered systems, starting with a Ni/Ag multilayer\cite{Tonnerre:740}, directly observed the antiferromagnetic peak of the artificial layered structure.  Similar studies were then made on Co/Cu\cite{Seve:68}, Fe/Mn\cite{Tonnerre:6293}, Fe/Co\cite{Sacchi:108} and Gd/Fe\cite{Hashizume:133} multilayer structures.   Two motivations inspired first multilayer soft x-ray reflectivity experiments.   Hase \emph{et al.}\cite{Hase:R3792,Hase:15331} looked at samples of Co/Cu multilayers with 25 and 50 repeats, showing that the magnetic interlayer roughness was one or two orders of magnitude larger than the structural roughness.   Sch\"{a}fers~\emph{et al.}\cite{Schafers:4074} used multilayer diffraction to  develop a soft x-ray polarimeter to detect the polarization state of soft x-ray beams.  Soft x-ray reflectivity measurements have facilitated a number of high profile papers on magnetic multilayer materials\cite{Durr:2166,Abes:07C703,Tonnerre:157202} that have potential impact in novel device manufacture.

Soft x-ray diffraction from single crystals was initially driven by the study of magnetic and orbital order in manganites, inspired by a theoretical paper by Castleton and Altarelli\cite{Castleton:1033} suggesting that the structural Jahn-Teller distortion and the orbital order (3d electric quadrupole moment) could be separated by diffraction at the $L_{2,3}$ edges.    After an initial study of the magnetic reflection in the antiferromagnet La$_{1.05}$Sr$_{1.95}$Mn$_2$O$_7$\cite{Wilkins:187201}, the orbital order was studied in a variety of layered manganites\cite{Wilkins:167205,Dhesi:056403,Thomas:237204,Grenier:206403,Wilkins:245102,Staub:214421,Wilkins:L323,HerreroMartin:224407,Beale:054433}, precipitating further theoretical analysis\cite{Stojic:104108,Mirone:23}.

Following these early results, soft x-ray diffraction has been used to study many more materials, not only transition metal oxides such as cuprates\cite{Abbamonte:155,Abbamonte:581,Fink:100502}, nickelates\cite{SchusslerLangeheine:156402,Scagnoli:100409} and cobaltates\cite{Chang:116401}, but has also been a powerful tool in deciphering the magnetic structure of complex rare earth materials\cite{Beale:174432,Mulders:11195,Mulders:184438}.   A recent, highly successful, experimental focus using soft x-ray diffraction has been to study multiferroic materials, including the $R$Mn$_2$O$_5$\cite{Koo:197601,Bodenthin:027201,Okamoto:157202} and $R$MnO$_3$\cite{Forrest:422205,Wilkins:207602} ($R$ = Rare earth), where the large magnetic structure and corresponding small wavevector lend themselves to soft x-ray diffraction.   A key result has been the ability to directly separate the components of the magnetic structure, something that is difficult either with neutron diffraction, where the bulk magnetic moment is seen, or with transition metal $K$ edge x-ray diffraction where the overlap between the higher lying unoccupied states that are probed severely complicates the analysis.   Although there are still many unanswered questions regarding the bulk oxides, soft x-ray diffraction has the potential to be a vital tool in the understanding\cite{Ohtomo:423} of interfaces between different systems where novel phenomena such as orbital reconstruction\cite{Chakhalian:1114}, and superconductivity between two insulating layers has been observed.   The relatively long wavelength of soft x-rays make it feasible to create a coherent beam with a small aperture in the beam prior to the sample.   This opens up a range of opportunities, including studies of magnetic speckle with coherent soft x-ray resonant magnetic scattering \cite{chesnel02,chesnel04} and soft x-ray resonant magnetic scattering of patterned samples \cite{dudzik,haznar, ogrin}.    

There are now a large number of dedicated soft x-ray diffractometers situated at synchrotron sources, many of which were influenced from a prototypical scattering vacuum chamber at Daresbury laboratory\cite{Roper:1101}.   Current instruments include the horizontal scattering RESOXS endstation at the SLS\cite{Staub:469} (Switzerland),  a 5-circle vertical diffraction chamber on the ID08 beamline at the ESRF\cite{Beutier:093901} (France), a horizontal scattering endstation at X1B\cite{Abbamonte:581} at the NSLS (USA),  reflectivity endstations on U4B and X13A also at the NSLS, and a horizontal scattering chamber on BL17SU\cite{Takeuchi:023905} at Spring8 (Japan), a diffraction and reflectivity endstation `ALICE' at BESSY\cite{Grabis:4048} (Germany), a scattering chamber at PLS (Korea) and a diffraction chamber on NSRRC (Taiwan).   In addition the authors are aware of instruments being commissioned at the Soleil Synchrotron (France), X1A at the NSLS, and 10ID-2 at the CLS (Canada).

\section{Instrument Description}

\subsection{Diffractometer}

The design for the RASOR diffractometer was based on the following criteria:
\begin{itemize}
\item{Small sphere of confusion}
\item{Excellent for both diffraction and reflectivity}
\item{Low temperature sample environment}
\item{Polarization analysis of scattered beam}
\item{Access to sample and polarization stages}
\item{Reliability}
\end{itemize}

\begin{figure}
\includegraphics[width=\columnwidth]{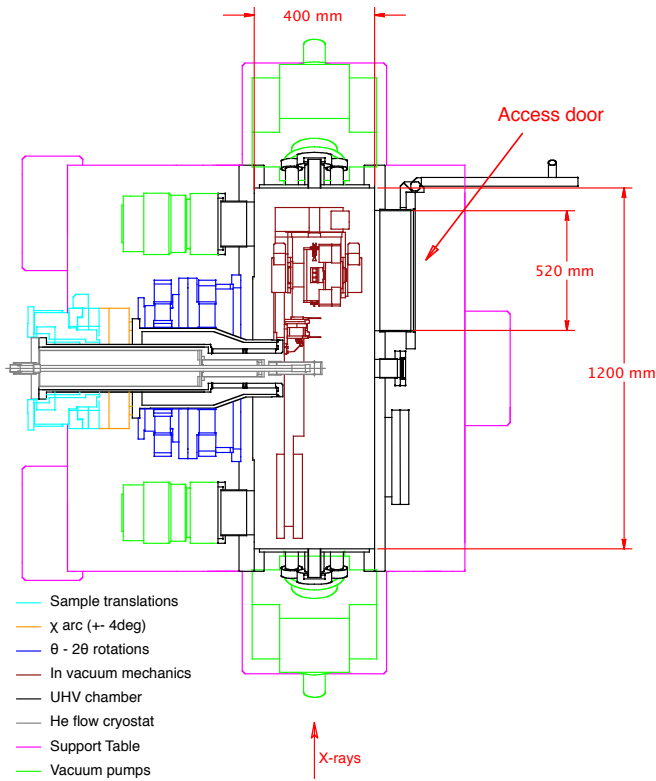}
\caption{\label{fig:plan}Plan diagram of the diffractometer.   The different components of the diffractometer are colour coded.   The internal mechanism includes two sets of apertures or slits between the sample and detector, and a complete polarization analysis mechanism.    The diagram is shown with the $2\theta$ arm at 0$^{\circ}$ when the detector would be in the main beam.    The dimensions specify the internal dimensions of the vacuum vessel.}
\end{figure}

A vertical scattering configuration was chosen to benefit from the low vertical divergence of the beam, particularly important in reflectivity measurements.   The rotation circles are all mounted on a single side of the vacuum chamber to minimise the effect of vacuum deformation on the sphere of confusion. RASOR was developed and manufactured from the conceptual design by Toyama Co. Ltd\cite{Toyama}.   The system comprises a ultra-high-vacuum (UHV) chamber with goniometer, mounted on a motorised table (Fig.~\ref{fig:plan}).    The goniometer provides $\theta$, $2\theta$, $\chi$ and sample translations through external stages.    Internal vacuum stages mounted on the $2\theta$ arm provide two sets of vertically defining apertures, and complete polarization analysis.   The sample is mounted directly onto the cold finger of a He$^4$ flow cryostat, that is itself mounted through the centre of the goniomoneter.

The table comprises of three motorised jack feet, upon which a translation of the diffractometer across the beam, and a rotation around a vertical axis is mounted.   These provide the necessary motions to position accurately the centre of rotation of the diffractometer onto the synchrotron beam.    A translation of the goniometer parallel to the x-ray beam is unnecessary due to the low divergence of the beam.

\begin{figure}
\includegraphics[width=\columnwidth]{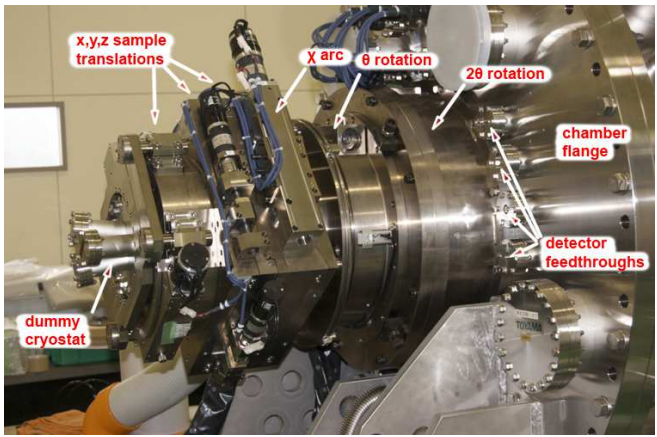}
\caption{\label{fig:ext}The external motion stages and back of the vacuum vessel.   Mounted on the chamber in the right of the photo, the $2\theta$, $\theta$, $\chi$ and sample translation stages can be seen from right to left.    A `dummy' cryostat is fitted for measuring the sphere of confusion.     The flange of the chamber is 50~mm thick and braced to minimise deflection from vacuum forces.    The vacuum pumps are not fitted for clarity.}
\end{figure}

The out-of-vacuum goniometer motions (Fig.~\ref{fig:ext}) provides full $\theta$ and $2\theta$ rotations through a set of double pumped differential seals.  High ratio gearboxes give a minimum step-size of 0.18~arcsec, and the position is controlled in closed loop with a Renishaw encoder with and accuracy of 1~arcsec.   A limited $\chi$ arc of $\pm\,4^{\circ}$ provides sufficient adjustment to align slightly mis-cut samples, and sample translations can adjust the position of the cryostat by $\pm\,5$~mm vertically and horizontally in the diffraction plane, and $\pm\,20$~mm perpendicular to the sample plane allowing different sample positions to be accessed.   The $\chi$ and sample translation motions are facilitated through two sets of bellows.    The sphere of confusion of the $\theta$, $2\theta$ and $\chi$ rotations is within 50~$\mu$m, even under vacuum.   

\begin{figure}
\includegraphics[width=\columnwidth]{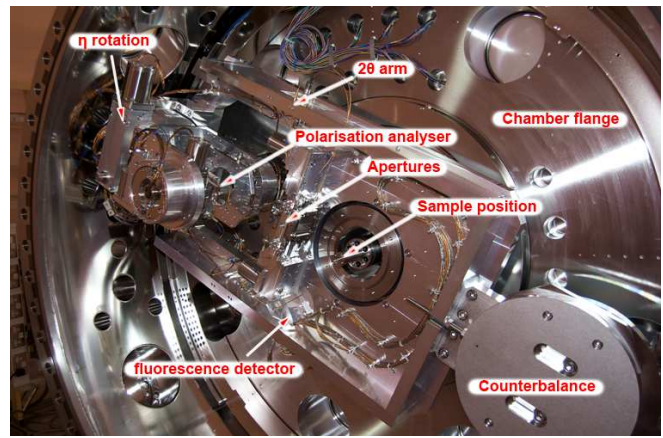}
\caption{\label{fig:int}The internal mechanisms mounted in the vacuum chamber.    The counterweight for the $2\theta$ arm can be seen at the bottom right, with the two sets of apertures and polarization analyser toward the top left.   A complex cable management system was designed to separately carry the motion and detector cables from the back flange of the chamber to the end of the $2\theta$ arm.}
\end{figure}

Two internal translation stages on the detector arm hold two aperture carriages 103~mm and 166~mm from the sample position (Fig.~\ref{fig:int}).   These hold a replaceable plate with vertically defining rectangular apertures.  The apertures are 20~mm wide horizontally, and have vertical dimensions of 50~$\mu$m, 100~$\mu$m, 500~$\mu$m, 1~mm, and 5~mm.    Further out on the 2$\theta$ arm is a complete polarization analyser.    Detectors for scattered radiation are mounted in line with the slit carriages on the polarization analyser, and also out of the diffraction plane for fluorescence and electron yield measurements (described below).

\subsubsection*{Polarization Analysis}

The diffractometer is equipped with sufficient motions for complete polarization analysis of the scattered beam\cite{Hill:140102}.   A second scattering process is conducted with the scattered beam from the sample forming the incident beam on the polarization analyser.    The analyser is selected such that the $2\theta$ angle for the x-ray energy is as close to 90$^{\circ}$ as possible.   The charge scattering is thus only sensitive to $\sigma$ (for the analyser) incident x-rays.    As the diffraction plane of the polarization analyser is rotated around the scattered beam ($\eta$), the intensity of the beam in each polarization state is measured.    Rather than measuring the intensity of just the $\sigma'$ and $\pi'$ polarization channels\cite{Staub:469,Staub:214421}, the error in the measurement can be dramatically reduced by measuring the intensity every few degrees of $\eta$ and then modelling the intensity with a sinusoidal function.  This has commonly been undertaken with soft x-ray polarimetry\cite{Schafers:4074}, however has not been performed in a soft x-ray diffraction setting.   Combining polarization analysis of the scattered beam with a rotation of the linear polarized incident x-rays, full linear polarization analysis\cite{Scagnoli:778,Mazzoli:195118,Johnson:104407} can be achieved.   

The very low x-ray energies force the use of multilayer materials rather than single crystals for use as an analyser.   Although this has disadvantages in the intensity of the reflected beam, the analysers can be designed such that the layer thickness produces a $2\theta$ diffraction angle that is much closer to 90$^{\circ}$ for each required energy than is possible with single crystals.

Multilayer crystals can either be grown as a graded crystal, such that a number of different energies can be optimised with one multilayer, or a series of multilayers with different spacings can be used.   RASOR is equipped with crossed translation stages on the $\theta$ analyser rotation such that either type can be accepted.   Furthermore, these translations allows the movement of the multilayer carriage out of the scattered beam.   A full $2\theta$ detector rotation on the polarization analyser allows the same detectors to be used for the direct beam and the polarization analyser.   The switch between these two configurations is possible under vacuum, assuming the required polarization analysers are mounted.   Multiple detectors can be mounted on the $2\theta$ rotation of the polarization analyser stage, such that a switching between detectors can be easily implemented under vacuum by redefining the zero point of this rotation.

\subsubsection*{Cryostat and sample environments}

The cryostat is a custom modified Janis SuperTran ST-400 continuous flow He$^4$ cryostat.   The modifications have the cryostat mounted on a custom flange using either a viton o-ring or helicaflex seal.    The cryostat has a rigid support tube that reduces in diameter to maximise the stiffness of the cryostat in the constricted space available.   Subsequent to fitting the cryostat the concentricity of the $\theta$ and $2\theta$ circles was adjusted to maintaining the sphere of confusion to $<50~\mu$m.   In addition to the standard feedthrough for the thermal control of the cryostat temperature, three more feedthroughs have been provided on the cryostat.    A 9-pin feedthrough is connected through kapton cable to a custom 9-pin male PEEK connector at the cold end on the cryostat.   This is easily accessible with the cryostat mounted through the access door, and a number of PEEK female connectors are provided for user experiments,  allowing regular setups to be configured.   Typically four of the nine pins will be used for a second temperature sensor mounted close to the sample, with the remaining 5 pins used either for a piezo rotation stage providing a $\phi$ rotation or in-situ transport measurements.    In addition to the 9 pin connector, there are two MHV feedthroughs on the cryostat connected though HV kapton cable to a second PEEK connector.   This has been specifically provided for applying an electric field to the sample, and each connector can support $\pm\,2$~kV.    Finally a triaxial BNC feedthough with a shielded coaxial kapton cable is provided for drain current measurements from the sample. 

\begin{figure}
\includegraphics[width=\columnwidth]{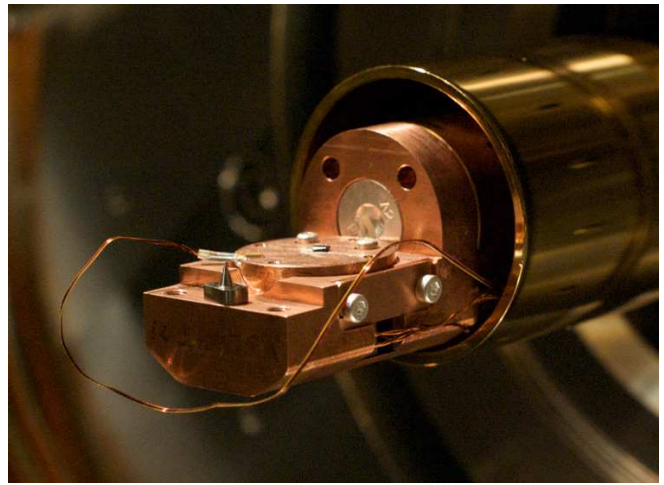}
\caption{\label{fig:atto}A sample mounted on the attocube piezo rotator providing a $\phi$ sample rotation.   The power and encoder cables for the attocube are visible feeding into the bottom of the sample mount   The wires in the foreground lead to a silicon diode temperature sensor on the top of the attocube.    The pin situated at the end of the sample mount is used for defining the centre of rotation, and aligning this onto the x-ray beam.  The gold sputtered short radiation shield allows maximum access to the sample position when the lowest sample temperature is not a priority.}
\end{figure}

Two radiation shields can be used with the cryostat.   These screw onto the cryostat and are cooled by the exhaust He$^4$ gas.   A short radiation shield (Fig.~\ref{fig:atto}) allows full access to the sample stage, and the sample can be mounted at any point on the sample stage and aligned on the beam with the sample translation motion.    A second radiation shield completely encloses the cold head of the cryostat except for a 10~mm wide slot covering 180$^\circ$ that is aligned such that the direct beam and any scattered beam can be detected.   The slot size is designed such that the full range of $\chi$ can still be utilised with this radiation shield.

The cryostat is terminated in a flat interface with a central tapped M6 hole surrounded by four 4/40 UNC tapped holes centred on a 19 mm diameter circle.   The central tapped hole provides the primary attachment for the sample mounts and the four surrounding holes provide rotational alignment.   Figure~\ref{fig:atto} shows a sample mount specifically designed for use with an attocube ANR50 piezo rotator.  The sample is mounted on closed cylinder sitting on the attocube.   The cylinder walls are machined such that a uniform 0.25~mm gap exists between the sample cylinder and the attocube mount.   Thermal conductance through the attocube has been shown to be extremely low, however with this configuration using the short radiation shield a base temperature of 80~K is easily achievable.   The attocube is positioned such that the slotted radiation shield can also be used to obtain lower temperatures.     The pin on the attocube mount (Fig.~\ref{fig:atto}) is used for optically determining the centre of rotation of the diffractometer, and positioning the centre of rotation in the x-ray beam.

A diffraction and reflectivity sample mount has also been designed for use without the attocube.   Using the short radiation shield a number of samples can be mounted for a single experiment, and a 40~mm translation of the cryostat along the $\theta$ axis enables different sample to be studied without breaking vacuum.   The radiations shields are attached to the cryostat and therefore translated with the cryostat.  As such when the slotted radiation shield is used, a sample can only be mounted in one position below the slot.  A base temperature of 15~K with the slotted radiation shield and 22~K with the short radiation shield has been achieved, however it is anticipated this will be improved by optimising the interface between the sample mount and the cryostat head.

The sample position is monitored using three cameras mounted on the top and front of the vacuum chamber and one facing the incident beam.   This viewport opposite the incident beam is fitted with a YAG crystal to show the beam position.    The centre of rotation can be determined using the top and front cameras while the diffractometer is under vacuum.   This centre of rotation can then be positioned onto the beam using the third camera.

\subsubsection*{Detector Electronics}

Two complementary detector systems have been installed.   For studies requiring a very large dynamic range, such as reflectivity measurements, or very strong signals, a 10 mm$^2$ UHV compatible IRD photodiode is used.   This can be used to detect the direct x-ray beam.   The small current from the photodiode is carried along the inner core of a secondary shielded coaxial cable to a floating SMA feedthrough.   This signal is then read by a Keithley 6514 Electrometer.   Currently the 0-2~V output from this is passed through a voltage to frequency converter and into a standard Diamond VME scalar, however the Electrometer can also be read directly from RS232 / GPIB.   The darkcurrent in the photodiode detector chain is in the region of 1.5~pA, with a noise of $\sim$0.1~pA.

For weak scattered signals, a KBr coated Burle 4869 channeltron${\textregistered}$ electron multiplier has been installed.   The KBr coating increases the efficiency of the detector to soft x-rays.   The channeltron${\textregistered}$ is driven with a potential difference of $\sim2~$kV.   In order to minimise the effect of detecting photoelectrons emitted from the sample, the front end of the channeltron${\textregistered}$ is charged to $-2~$kV, with the other end connected to ground.   The signal is carried to an SMA feedthrough by a kapton coaxial cable, and is then amplified by an Ortec VT120 fast amplifier mounted on the chamber next to the feedthrough.   The amplified pulses are then shaped by a FAST ComTec constant fraction discriminator that outputs TTL pulses  to the scalar.   The background counts from the channeltron$\textregistered$ is in the region of 0.1~s$^{-1}$.

Another 10~mm$^2$ photodiode and 4869 channeltron${\textregistered}$ detector are mounted slightly out of the diffraction plane, 208 mm and 165 mm respectively from the sample position.   These can be used for fluorescence and electron yield measurements.   The detector electronics are identical to that described above, with the exception of a reverse polarity on the channeltron${\textregistered}$ to optimise the electron detection.

\subsubsection*{Control System}

RASOR is controlled through the Diamond Light Source EPICS\cite{epics} system, and all vacuum and motion equipment is controlled directly through EPICS drivers.   User experiments are conducted through GDA\cite{gda} which provided a user friendly graphical environment for data acquisition.   This allows the user to scan both physical and pseudo motors, while counters can be defined as any EPICS process variable.

\subsubsection*{Vacuum System}

The RASOR diffractometer is primarily designed as a user instrument.   Therefore the vacuum system is designed to be both simple to use and fully interlocked.    The primary pumping is provided by two 400~l.s$^{-1}$ ion pumps, and  two 500~l.s$^{-1}$ turbo molecular pumps.    The ion pumps are positioned either side of the chamber, one below the incident beam port, and the other symmetrically opposite.  The two 500~l.s$^{-1}$ turbo molecular pumps mounted on the back flange below the differential feedthrough.   The ion pumps are equipped with titanium sublimation NEG pumps, with a cryo shroud cooled with liquid nitrogen on one pump and an ambient shroud on the other pump.

Prior to baking the system was able to achieve vacuum in the region of $3\times10^{-8}~$mbar, and no presence of ice was observed while maintaining the sample at 15 K for 12 hours.    The $\theta$ and $2\theta$ rotation motions are provided by a double pumped differential feedthrough.   This has two intermediary stages, the first high vacuum chamber separated from the experimental vacuum by a viton o-ring and pumped with an 80~l.s$^{-1}$ turbo molecular pump.   The second low vacuum stage, isolated through a second viton seal from the high vacuum stage, and a third viton seal from atmosphere is pumped by a scroll pump that also backs the turbo pump for the first stage.

Initial vacuum in the system is provides by a roots pump, that achieves a pressure of $10^{-2}$~mbar in the chamber in 20 minutes.    A further 2 hours of pumping with the two 500~l.s$^{-1}$ turbo molecular pumps is required before the chamber can be opened to the x-ray beam, and 8 hours of pumping is required before the sample can be cooled to 15~K, to ensure no ice forms on the sample surface.

\subsection{Beamline and x-ray source characteristics}

The RASOR endstation is available for general user proposals, and currently is accommodated on the I06-I nanoscience branchline at Diamond Light Source.    This beamline is equipped with an APPLE-II undulator capable of providing right and left circularly polarized light, and linearly polarized light in the energy range 80 to 2100~eV, with an energy resolving power of $10^4$ at 400~eV.   In addition to horizontal and vertical linear polarizations, the undulator has been commissioned to provide a rotation of the linear polarized light over 90$^{\circ}$.   This allows measurements of the anisotropy of the scattering process and combined with the polarization analyser provides full linear polarization analysis.   The branchline beam is focussed using a 2:1 mirror providing a beamsize of 20~$\mu$ vertically and 200~$\mu$m horizontally.   The smaller beamsize reduces the overall beam intensity, and this can be adjusted with the focussing mirror entrance slits.

RASOR will ultimately become a permanent endstation on beamline I10 (BLADE), also at Diamond, which is currently in development.   User experiments with RASOR on I10 are scheduled for 2011.    Beamline I10 will have a number of facilities not available on I06-I, including fast (10~Hz) switching of the beam polarization state, and a rotation of the linear polarization through 180$^\circ$.

\section{Experimental Results}

\subsection{Reflectivity}

\begin{figure}
\includegraphics[width=\columnwidth]{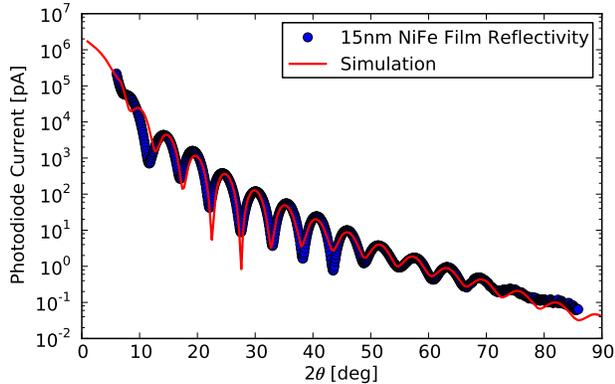}
\caption{\label{fig:ref}Reflectivity spectra from a 15~nm film of NiFe alloy, taken at 845~eV off resonant below the Ni $L$ edges.   Fringes can clearly be seen out to 80$^{\circ}$ in $2\theta$, and the data spans six orders of magnitude of intensity with a noise level of $<0.1~$pA.  The solid red line shows the simulated reflectivity.}
\end{figure}

Reflectivity spectra were taken from 10 and 15~nm films of NiFe on an silicon substrate.   The films were magnetised with a NdFeB magnet parallel with the incident beam direction prior to mounting in the diffraction chamber.    Figure~\ref{fig:ref} shows the specular reflectivity measured off resonance at 845~eV just below the nickel $L$ edges.    The spectra shows fringes visible to $>80^{\circ}$, spanning 6 orders of magnitude.   The data was measured with a 500~$\mu$m slit in front of the detector, and a 1~mm scatter slit closer to the sample.   The positioning of the sample on the centre of rotation and the incidence of the x-ray beam on this point is critical to the measurement of high angle reflectivity data.    Prior to the measurements the centre of rotation was defined optically using a fine pin mounted adjacent to the sample, and the position of the diffractometer was adjusted such that the centre of rotation was accurately positioned within the beam.    The specular reflectivity was simulated using the GenX program\cite{bjorck:1174} that uses the Parratt recursion formula\cite{Parratt:359}.   The best simulation of the data was generated using a 16.5~nm film of NiFe, with an additional 1~nm thick film with a slightly lower density above the NiFe layer.   The numerical fitting returned a roughness between the NiFe and the silicon substrate, and at the top of the NiFe layer of $\sim$3~nm.  Although the information retrieved from this simple sample is relatively trivial, the experiment demonstrates the ability of the diffractometer to measure Kiessig fringes to a very high $\vec{Q}$.   This is due to a combination of precisely encoded $\theta$ and $2\theta$ stages, and the low divergence of the incident x-ray beam.   The ability to accurately position the sample onto the centre of rotation, and the coincidence of the centre of rotation with the incident beam prevents the measurement from deviating from the specular ridge.

\begin{figure}
\includegraphics[width=\columnwidth]{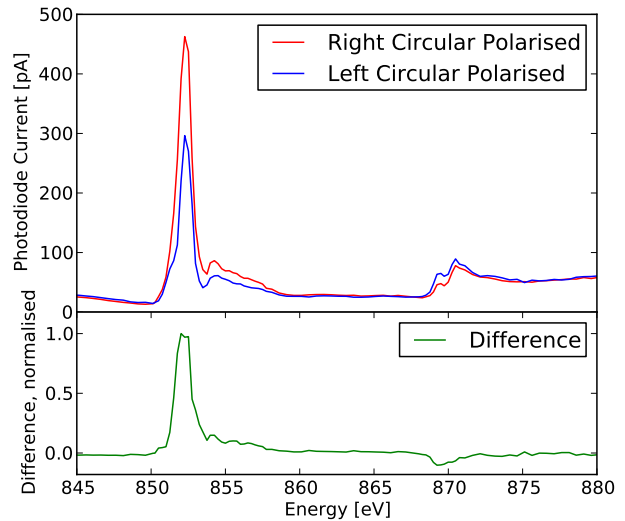}
\caption{\label{fig:dic}Dichroism of a magnetised 10~nm NiFe film at the Ni $L$ edges.   The lower panel shows the difference between the resonant intensity with left and right circularly polarized incident x-rays.   The data was collected at room temperature with the photodiode detector.}
\end{figure}

The absorption of x-rays by a ferromagnetic material depends on the polarization state of the incident x-rays, giving a dichroism response\cite{Schutz:737}.   Although this is normally done by measuring the absorption, similar information can be retrieved from the energy dependence of the reflectivity.   Figure~\ref{fig:dic} shows the energy dependence of the specular reflectivity of a 10~nm NiFe film at $2\theta=30^{\circ}$ with left and right circularly polarized x-rays, with the lower panel showing the difference.   

\subsection{Diffraction}

\subsubsection*{Magnetic (001) reflection in \lsmo}

\begin{figure}
\includegraphics[width=\columnwidth]{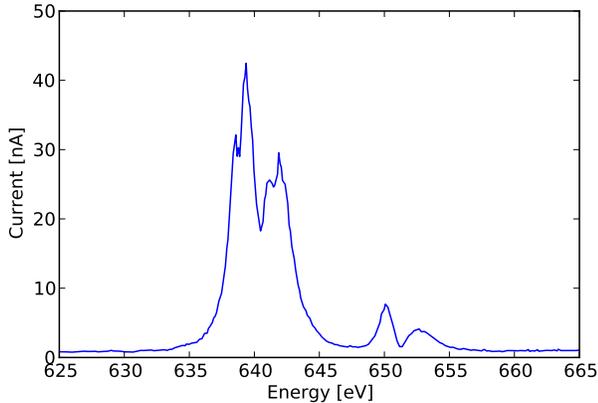}
\caption{\label{fig:mn}Resonance of the (001) magnetic reflection in \lsmo.   Data was collected with a photodiode detector with the sample at 15~K.   Previously unobserved structure in the $L_3$ edge at 635-643~eV can be seen.}
\end{figure}

\lsmo\ is a bilayer manganite formed by layers of MnO$_6$ octahedra separated by (La,Sr)O layers.   This forms a very two-dimensional system, with phenomena such as charge and orbital ordering occurring within the MnO$_6$ layers that form the $ab$ plane.   Below T$_N\approx180~$K, these layers form ferromagnetic sheets which are then coupled antiferromagnetically through superexchange between the sheets.    Due to the $I4/mmm$ space group the (001) Bragg reflection is normally forbidden, however the magnetic structure breaks the symmetry and a magnetic (001) reflection is allowed.   This has previously been observed using soft x-ray diffraction\cite{Wilkins:187201}.     Due to the very long (19.95~\AA) $c$-axis, the (002) reflection of \lsmo\ is within the Ewald sphere at the Mn $L$ edge.     Figure~\ref{fig:mn} shows the resonant enhancement of the (001) magnetic reflection through the Mn $L_3$ (635-645~eV) and $L_2$ (648-658~eV) edges.   By comparison to previous results, the resonance shows significantly more structure at the $L_3$ edge.   Whereas previous diffraction data had shown two main peaks at 638~eV and 642~eV, at least three distinct features can now be seen within each of these peaks.   This is an indication of both the high energy resolution of I06 and the high angular resolution of RASOR.

\begin{figure}
\includegraphics[width=\columnwidth]{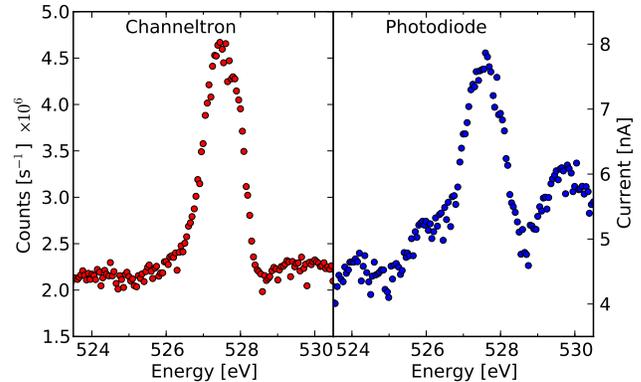}
\caption{\label{fig:o}Oxygen resonance of the magnetic (001) reflection in \lsmo.   The resonance is speculated to arise from hybridisation between the Mn and O in the MnO$_6$ octahedra.     The resonance was measured at 15 K with both the photodiode and channeltron$\textregistered$ detectors.}
\end{figure}

In addition to resonance at the Mn $L_3$ edge, a resonant signal was observed at the oxygen $K$ edge (Fig.~\ref{fig:o}).   There are two possible explananations for the presence of this signal.   Either there is a small magnetic moment on the oxygen ions induced by a hybridization between the manganese ions and oxygen ligands, or alternatively the signal arises from an anisotropic scattering factor due to an anisotropic oxygen $2p$ orbital.   This anisotropy breaks the global symmetry of the $I4/mmm$ structure and thus a weak signal is seen at the otherwise forbidden (001) Bragg position.  

The weak resonance at the oxygen edge, much weaker than that at the manganese edge, was used as a test for the two detector systems.   Figure~\ref{fig:o} shows the resonance measured both with the photodiode and channeltron$\textregistered$ detector.   It is clear with this reflection that the channeltron$\textregistered$ shows a much higher signal to noise ratio, and it is presumed that for much weaker signals this will be of even greater benefit.   As the channeltron$\textregistered$ is a photon rather than a photon flux detector the gain in the signal to noise ratio through increased counting times is much more pronounced.

\subsubsection*{Incident polarization dependence of charge scattering in LuFe$_2$O$_4$}

The mechanism enabling multiferroic properties in materials is far from fully understood, however it is common for materials to possess either a cycloidal magnetic order or occasionally antiferromagnetic order that interacts with an electric polarization from the covalent bonding between anions.   In contrast, the electric polarization in LuFe$_2$O$_4$ may well be driven by a frustrated charge order.   Thus the study of this charge order is particularly important not only to understand the properties of LuFe$_2$O$_4$, but of multiferroicity in general.    This material has the additional benefit in that these phenomena occur at room temperature, and thus are potentially useful for novel device development.

\begin{figure}
\includegraphics[width=\columnwidth]{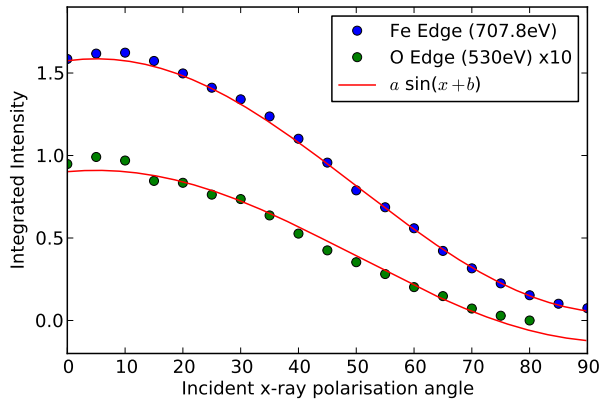}
\caption{\label{fig:pol}Intensity of the $(2\tau, -\tau, 3/2)$ charge order reflection at the Fe $L_2$ edge as a function of the angle of the incident beam polarization.   The data was measured using the photodiode detector chain with a sample at 285~K.  The dependence fits a sinusoidal dependance, }
\end{figure}

The charge order in LuFe$_2$O$_4$ has been observed\cite{Mulders:077602} through reflections at both (1/3,1/3,$n$/2) where $n=$odd and (2$\tau$,$\tau$,$n$) type positions.   Here we observe the charge order reflection at (2$\tau$,$-\tau$,3/2) where $\tau=0.028$ at the Fe $L_3$ edge and oxygen $K$ edge.   A particularly useful aspect of resonant scattering is the ability to determine the anisotropy of the diffraction signal which can uncover both the anisotropy of the system and help determine the origin of the scattering.    Fig.~\ref{fig:pol} demonstrates the ability to observe such anisotropy with soft x-ray diffraction.   The angle of linear polarization of the x-ray beam was altered in 5$^{\circ}$ steps from 0$^{\circ}$, equivalent to horizontal linearly polarized x-rays, to 90$^{\circ}$, corresponding to vertical polarization.    The highest intensity was observed with close to horizontal linear polarization with an $a\sin (x+b)$ function closely fitting the data.   The high intensity with horizontally polarized (or $\sigma$-incident) x-rays is expected for a charge order reflection, however more detailed analysis and further data will be reported in a future publication\cite{bland}.

\section{Upgrade path}

The diffractometer is specifically designed such that upgrades can easily be encompassed, tailored either to a general user requirement, or for an individual experiment.    Of particular interest and demand is an applied magnetic field for the sample.    Suitable mounting holes have been provided on the internal $\theta$ rotation, and two magnets are currently in development.   Firstly, a dipole electromagnet aligned with the field direction along the beam, producing a field in the order of 0.1~T.    Secondly a more complex vector field magnet that will  enable the complete control of the magnetic field direction within the plane of the sample surface.    Finally there is the provision for a much larger ($\sim$1~T) electromagnet in a fixed position.   All of these magnets could be used with the $\phi$ sample rotation to investigate any anisotropy of the effect caused by an applied magnetic field.

In addition to mounting points on the $\theta$ circle, the $2\theta$ arm has been specifically designed to accept an addition 5~kg load at the detector position without compromising the sphere of confusion.   This is primary to accommodate a 2D area detector.    There are a number of options for area detectors, however designing a UHV system reduces the options.   There is a commercially available CCD based detector that has been successfully used at the Swiss Light Source.   Potentially of greater interest is the development of CMOS detectors.   These can be designed with a much higher readout time, and a higher dynamic range than CCD chips.   In particular the ``Vanilla'' detector chip\cite{Blue:287} that has recently been characterized running in a back-thinned mode for ultra-violet detection\cite{Blue:215}, developed through the M-I$^3$ consortium\cite{MI3} may well be applicable both for soft x-ray and UHV conditions.

\section{Conclusion}

We have developed a soft x-ray diffractometer for single crystal diffraction and thin film reflectivity, and installed the instrument at Diamond Light Source.   We have demonstrated the capability of the instrument through reflectivity and dichroism measurements from NiFe thin films, and single crystal diffraction from LuFe$_2$O$_4$ and \lsmo.     The potential for full polarization analysis by combining a rotation of the incident linear beam polarization has been described, and the rotation of the incident beam demonstrated through measurements of anisotropic charge reflections.

\section*{Acknowledgements}

The authors would like to thank B.K. Tanner and S.B. Wilkins for advice throughout the project, and S.R. Bland, R.D. Johnson, M.S. Brewer, M. Sussmuth, S. Cavill and C. Richardson for experimental assistance.   The project was funded through a CLRC (STFC) facility development grant.  We would also like to thank Diamond Light Source and EPSRC for financial support.


\begin{thebibliography}{73}
\expandafter\ifx\csname natexlab\endcsname\relax\def\natexlab#1{#1}\fi
\expandafter\ifx\csname bibnamefont\endcsname\relax
  \def\bibnamefont#1{#1}\fi
\expandafter\ifx\csname bibfnamefont\endcsname\relax
  \def\bibfnamefont#1{#1}\fi
\expandafter\ifx\csname citenamefont\endcsname\relax
  \def\citenamefont#1{#1}\fi
\expandafter\ifx\csname url\endcsname\relax
  \def\url#1{\texttt{#1}}\fi
\expandafter\ifx\csname urlprefix\endcsname\relax\def\urlprefix{URL }\fi
\providecommand{\bibinfo}[2]{#2}
\providecommand{\eprint}[2][]{\url{#2}}

\bibitem[{\citenamefont{van~der Laan}(2006)}]{Laan:120}
\bibinfo{author}{\bibfnamefont{G.}~\bibnamefont{van~der Laan}},
  \bibinfo{journal}{Current Opinion in Solid State and Materials Science}
  \textbf{\bibinfo{volume}{10}}, \bibinfo{pages}{120} (\bibinfo{year}{2006}).

\bibitem[{\citenamefont{van~der Laan}(2008)}]{Laan:570}
\bibinfo{author}{\bibfnamefont{G.}~\bibnamefont{van~der Laan}},
  \bibinfo{journal}{Comptes Rendus Physique} \textbf{\bibinfo{volume}{9}},
  \bibinfo{pages}{570} (\bibinfo{year}{2008}).

\bibitem[{\citenamefont{Spencer et~al.}(2005)\citenamefont{Spencer, Wilkins,
  Hatton, Brown, Hase, Purton, and Fort}}]{Spencer:1725}
\bibinfo{author}{\bibfnamefont{P.~D.} \bibnamefont{Spencer}},
  \bibinfo{author}{\bibfnamefont{S.~B.} \bibnamefont{Wilkins}},
  \bibinfo{author}{\bibfnamefont{P.~D.} \bibnamefont{Hatton}},
  \bibinfo{author}{\bibfnamefont{S.~D.} \bibnamefont{Brown}},
  \bibinfo{author}{\bibfnamefont{T.~P.~A.} \bibnamefont{Hase}},
  \bibinfo{author}{\bibfnamefont{J.~A.} \bibnamefont{Purton}},
  \bibnamefont{and} \bibinfo{author}{\bibfnamefont{D.}~\bibnamefont{Fort}},
  \bibinfo{journal}{Journal of Physics: Condensed Matter}
  \textbf{\bibinfo{volume}{17}}, \bibinfo{pages}{1725} (\bibinfo{year}{2005}).

\bibitem[{\citenamefont{Jark and St\"{o}hr}(1988)}]{Jark:654}
\bibinfo{author}{\bibfnamefont{W.}~\bibnamefont{Jark}} \bibnamefont{and}
  \bibinfo{author}{\bibfnamefont{J.}~\bibnamefont{St\"{o}hr}},
  \bibinfo{journal}{Nuclear Instruments and Methods in Physics Research Section
  A: Accelerators, Spectrometers, Detectors and Associated Equipment}
  \textbf{\bibinfo{volume}{266}}, \bibinfo{pages}{654} (\bibinfo{year}{1988}).

\bibitem[{\citenamefont{Kao et~al.}(1990)\citenamefont{Kao, Hastings, Johnson,
  Siddons, Smith, and Prinz}}]{kao:373}
\bibinfo{author}{\bibfnamefont{C.}~\bibnamefont{Kao}},
  \bibinfo{author}{\bibfnamefont{J.~B.} \bibnamefont{Hastings}},
  \bibinfo{author}{\bibfnamefont{E.~D.} \bibnamefont{Johnson}},
  \bibinfo{author}{\bibfnamefont{D.~P.} \bibnamefont{Siddons}},
  \bibinfo{author}{\bibfnamefont{G.~C.} \bibnamefont{Smith}}, \bibnamefont{and}
  \bibinfo{author}{\bibfnamefont{G.~A.} \bibnamefont{Prinz}},
  \bibinfo{journal}{Phys. Rev. Lett.} \textbf{\bibinfo{volume}{65}},
  \bibinfo{pages}{373} (\bibinfo{year}{1990}).

\bibitem[{\citenamefont{Hannon et~al.}(1988)\citenamefont{Hannon, Trammell,
  Blume, and Gibbs}}]{Hannon:1245}
\bibinfo{author}{\bibfnamefont{J.~P.} \bibnamefont{Hannon}},
  \bibinfo{author}{\bibfnamefont{G.~T.} \bibnamefont{Trammell}},
  \bibinfo{author}{\bibfnamefont{M.}~\bibnamefont{Blume}}, \bibnamefont{and}
  \bibinfo{author}{\bibfnamefont{D.}~\bibnamefont{Gibbs}},
  \bibinfo{journal}{Phys. Rev. Lett.} \textbf{\bibinfo{volume}{61}},
  \bibinfo{pages}{1245} (\bibinfo{year}{1988}).

\bibitem[{\citenamefont{Kao et~al.}(1994)\citenamefont{Kao, Chen, Johnson,
  Hastings, Lin, Ho, Meigs, Brot, Hulbert, Idzerda et~al.}}]{Kao:9599}
\bibinfo{author}{\bibfnamefont{C.-C.} \bibnamefont{Kao}},
  \bibinfo{author}{\bibfnamefont{C.~T.} \bibnamefont{Chen}},
  \bibinfo{author}{\bibfnamefont{E.~D.} \bibnamefont{Johnson}},
  \bibinfo{author}{\bibfnamefont{J.~B.} \bibnamefont{Hastings}},
  \bibinfo{author}{\bibfnamefont{H.~J.} \bibnamefont{Lin}},
  \bibinfo{author}{\bibfnamefont{G.~H.} \bibnamefont{Ho}},
  \bibinfo{author}{\bibfnamefont{G.}~\bibnamefont{Meigs}},
  \bibinfo{author}{\bibfnamefont{J.-M.} \bibnamefont{Brot}},
  \bibinfo{author}{\bibfnamefont{S.~L.} \bibnamefont{Hulbert}},
  \bibinfo{author}{\bibfnamefont{Y.~U.} \bibnamefont{Idzerda}},
  \bibnamefont{et~al.}, \bibinfo{journal}{Phys Rev B}
  \textbf{\bibinfo{volume}{50}}, \bibinfo{pages}{9599} (\bibinfo{year}{1994}).

\bibitem[{\citenamefont{Chen and Sette}(1990)}]{Chen:119}
\bibinfo{author}{\bibfnamefont{C.}~\bibnamefont{Chen}} \bibnamefont{and}
  \bibinfo{author}{\bibfnamefont{F.}~\bibnamefont{Sette}},
  \bibinfo{journal}{Phys Scripta} \textbf{\bibinfo{volume}{T31}},
  \bibinfo{pages}{119} (\bibinfo{year}{1990}).

\bibitem[{\citenamefont{Tonnerre et~al.}(1995)\citenamefont{Tonnerre, S\`eve,
  Raoux, Soulli\'e, Rodmacq, and Wolfers}}]{Tonnerre:740}
\bibinfo{author}{\bibfnamefont{J.~M.} \bibnamefont{Tonnerre}},
  \bibinfo{author}{\bibfnamefont{L.}~\bibnamefont{S\`eve}},
  \bibinfo{author}{\bibfnamefont{D.}~\bibnamefont{Raoux}},
  \bibinfo{author}{\bibfnamefont{G.}~\bibnamefont{Soulli\'e}},
  \bibinfo{author}{\bibfnamefont{B.}~\bibnamefont{Rodmacq}}, \bibnamefont{and}
  \bibinfo{author}{\bibfnamefont{P.}~\bibnamefont{Wolfers}},
  \bibinfo{journal}{Phys. Rev. Lett.} \textbf{\bibinfo{volume}{75}},
  \bibinfo{pages}{740} (\bibinfo{year}{1995}).

\bibitem[{\citenamefont{S{\`e}ve et~al.}(1995)\citenamefont{S{\`e}ve, Tonnerre,
  Raoux, Bobo, Piecuch, Santis, Troussel, Brot, Chakarian, Kao
  et~al.}}]{Seve:68}
\bibinfo{author}{\bibfnamefont{L.}~\bibnamefont{S{\`e}ve}},
  \bibinfo{author}{\bibfnamefont{J.}~\bibnamefont{Tonnerre}},
  \bibinfo{author}{\bibfnamefont{D.}~\bibnamefont{Raoux}},
  \bibinfo{author}{\bibfnamefont{J.}~\bibnamefont{Bobo}},
  \bibinfo{author}{\bibfnamefont{M.}~\bibnamefont{Piecuch}},
  \bibinfo{author}{\bibfnamefont{M.~D.} \bibnamefont{Santis}},
  \bibinfo{author}{\bibfnamefont{P.}~\bibnamefont{Troussel}},
  \bibinfo{author}{\bibfnamefont{J.}~\bibnamefont{Brot}},
  \bibinfo{author}{\bibfnamefont{V.}~\bibnamefont{Chakarian}},
  \bibinfo{author}{\bibfnamefont{C.}~\bibnamefont{Kao}}, \bibnamefont{et~al.},
  \bibinfo{journal}{Journal of Magnetism and Magnetic Materials}
  \textbf{\bibinfo{volume}{148}}, \bibinfo{pages}{68} (\bibinfo{year}{1995}).

\bibitem[{\citenamefont{Tonnerre et~al.}(1998)\citenamefont{Tonnerre, Seve,
  Barbara-Dechelette, Bartolome, Raoux, Chakarian, Kao, Fischer, Andrieu, and
  Fruchart}}]{Tonnerre:6293}
\bibinfo{author}{\bibfnamefont{J.~M.} \bibnamefont{Tonnerre}},
  \bibinfo{author}{\bibfnamefont{L.}~\bibnamefont{Seve}},
  \bibinfo{author}{\bibfnamefont{A.}~\bibnamefont{Barbara-Dechelette}},
  \bibinfo{author}{\bibfnamefont{F.}~\bibnamefont{Bartolome}},
  \bibinfo{author}{\bibfnamefont{D.}~\bibnamefont{Raoux}},
  \bibinfo{author}{\bibfnamefont{V.}~\bibnamefont{Chakarian}},
  \bibinfo{author}{\bibfnamefont{C.~C.} \bibnamefont{Kao}},
  \bibinfo{author}{\bibfnamefont{H.}~\bibnamefont{Fischer}},
  \bibinfo{author}{\bibfnamefont{S.}~\bibnamefont{Andrieu}}, \bibnamefont{and}
  \bibinfo{author}{\bibfnamefont{O.}~\bibnamefont{Fruchart}},
  \bibinfo{journal}{Journal of Applied Physics} \textbf{\bibinfo{volume}{83}},
  \bibinfo{pages}{6293} (\bibinfo{year}{1998}).

\bibitem[{\citenamefont{Sacchi et~al.}(1998)\citenamefont{Sacchi, Hague,
  Gullikson, and Underwood}}]{Sacchi:108}
\bibinfo{author}{\bibfnamefont{M.}~\bibnamefont{Sacchi}},
  \bibinfo{author}{\bibfnamefont{C.~F.} \bibnamefont{Hague}},
  \bibinfo{author}{\bibfnamefont{E.~M.} \bibnamefont{Gullikson}},
  \bibnamefont{and} \bibinfo{author}{\bibfnamefont{J.~H.}
  \bibnamefont{Underwood}}, \bibinfo{journal}{Phys. Rev. B}
  \textbf{\bibinfo{volume}{57}}, \bibinfo{pages}{108} (\bibinfo{year}{1998}).

\bibitem[{\citenamefont{Hashizume et~al.}(1998)\citenamefont{Hashizume,
  Ishimatsu, Sakata, Iizuka, Hosoito, Namikawa, Iwazumi, Srajer, Venkataraman,
  Lang et~al.}}]{Hashizume:133}
\bibinfo{author}{\bibfnamefont{H.}~\bibnamefont{Hashizume}},
  \bibinfo{author}{\bibfnamefont{N.}~\bibnamefont{Ishimatsu}},
  \bibinfo{author}{\bibfnamefont{O.}~\bibnamefont{Sakata}},
  \bibinfo{author}{\bibfnamefont{T.}~\bibnamefont{Iizuka}},
  \bibinfo{author}{\bibfnamefont{N.}~\bibnamefont{Hosoito}},
  \bibinfo{author}{\bibfnamefont{K.}~\bibnamefont{Namikawa}},
  \bibinfo{author}{\bibfnamefont{T.}~\bibnamefont{Iwazumi}},
  \bibinfo{author}{\bibfnamefont{G.}~\bibnamefont{Srajer}},
  \bibinfo{author}{\bibfnamefont{C.}~\bibnamefont{Venkataraman}},
  \bibinfo{author}{\bibfnamefont{J.}~\bibnamefont{Lang}}, \bibnamefont{et~al.},
  \bibinfo{journal}{Physica B: Condensed Matter}
  \textbf{\bibinfo{volume}{248}}, \bibinfo{pages}{133} (\bibinfo{year}{1998}).

\bibitem[{\citenamefont{Hase et~al.}(2000{\natexlab{a}})\citenamefont{Hase,
  Pape, Tanner, D\"urr, Dudzik, van~der Laan, Marrows, and
  Hickey}}]{Hase:R3792}
\bibinfo{author}{\bibfnamefont{T.~P.~A.} \bibnamefont{Hase}},
  \bibinfo{author}{\bibfnamefont{I.}~\bibnamefont{Pape}},
  \bibinfo{author}{\bibfnamefont{B.~K.} \bibnamefont{Tanner}},
  \bibinfo{author}{\bibfnamefont{H.}~\bibnamefont{D\"urr}},
  \bibinfo{author}{\bibfnamefont{E.}~\bibnamefont{Dudzik}},
  \bibinfo{author}{\bibfnamefont{G.}~\bibnamefont{van~der Laan}},
  \bibinfo{author}{\bibfnamefont{C.~H.} \bibnamefont{Marrows}},
  \bibnamefont{and} \bibinfo{author}{\bibfnamefont{B.~J.}
  \bibnamefont{Hickey}}, \bibinfo{journal}{Phys. Rev. B}
  \textbf{\bibinfo{volume}{61}}, \bibinfo{pages}{R3792}
  (\bibinfo{year}{2000}{\natexlab{a}}).

\bibitem[{\citenamefont{Hase et~al.}(2000{\natexlab{b}})\citenamefont{Hase,
  Pape, Read, Tanner, D\"urr, Dudzik, van~der Laan, Marrows, and
  Hickey}}]{Hase:15331}
\bibinfo{author}{\bibfnamefont{T.~P.~A.} \bibnamefont{Hase}},
  \bibinfo{author}{\bibfnamefont{I.}~\bibnamefont{Pape}},
  \bibinfo{author}{\bibfnamefont{D.~E.} \bibnamefont{Read}},
  \bibinfo{author}{\bibfnamefont{B.~K.} \bibnamefont{Tanner}},
  \bibinfo{author}{\bibfnamefont{H.}~\bibnamefont{D\"urr}},
  \bibinfo{author}{\bibfnamefont{E.}~\bibnamefont{Dudzik}},
  \bibinfo{author}{\bibfnamefont{G.}~\bibnamefont{van~der Laan}},
  \bibinfo{author}{\bibfnamefont{C.~H.} \bibnamefont{Marrows}},
  \bibnamefont{and} \bibinfo{author}{\bibfnamefont{B.~J.}
  \bibnamefont{Hickey}}, \bibinfo{journal}{Phys. Rev. B}
  \textbf{\bibinfo{volume}{61}}, \bibinfo{pages}{15331}
  (\bibinfo{year}{2000}{\natexlab{b}}).

\bibitem[{\citenamefont{Sch\"{a}fers et~al.}(1999)\citenamefont{Sch\"{a}fers,
  Mertins, Gaupp, Gudat, Mertin, Packe, Schmolla, Fonzo, Soulli\'{e}, Jark
  et~al.}}]{Schafers:4074}
\bibinfo{author}{\bibfnamefont{F.}~\bibnamefont{Sch\"{a}fers}},
  \bibinfo{author}{\bibfnamefont{H.-C.} \bibnamefont{Mertins}},
  \bibinfo{author}{\bibfnamefont{A.}~\bibnamefont{Gaupp}},
  \bibinfo{author}{\bibfnamefont{W.}~\bibnamefont{Gudat}},
  \bibinfo{author}{\bibfnamefont{M.}~\bibnamefont{Mertin}},
  \bibinfo{author}{\bibfnamefont{I.}~\bibnamefont{Packe}},
  \bibinfo{author}{\bibfnamefont{F.}~\bibnamefont{Schmolla}},
  \bibinfo{author}{\bibfnamefont{S.~D.} \bibnamefont{Fonzo}},
  \bibinfo{author}{\bibfnamefont{G.}~\bibnamefont{Soulli\'{e}}},
  \bibinfo{author}{\bibfnamefont{W.}~\bibnamefont{Jark}}, \bibnamefont{et~al.},
  \bibinfo{journal}{Appl. Opt.} \textbf{\bibinfo{volume}{38}},
  \bibinfo{pages}{4074} (\bibinfo{year}{1999}).

\bibitem[{\citenamefont{D\"{u}rr et~al.}(1999)\citenamefont{D\"{u}rr, Dudzik,
  Dhesi, Goedkoop, van~der Laan, Belakhovsky, Mocuta, Marty, and
  Samson}}]{Durr:2166}
\bibinfo{author}{\bibfnamefont{H.~A.} \bibnamefont{D\"{u}rr}},
  \bibinfo{author}{\bibfnamefont{E.}~\bibnamefont{Dudzik}},
  \bibinfo{author}{\bibfnamefont{S.~S.} \bibnamefont{Dhesi}},
  \bibinfo{author}{\bibfnamefont{J.~B.} \bibnamefont{Goedkoop}},
  \bibinfo{author}{\bibfnamefont{G.}~\bibnamefont{van~der Laan}},
  \bibinfo{author}{\bibfnamefont{M.}~\bibnamefont{Belakhovsky}},
  \bibinfo{author}{\bibfnamefont{C.}~\bibnamefont{Mocuta}},
  \bibinfo{author}{\bibfnamefont{A.}~\bibnamefont{Marty}}, \bibnamefont{and}
  \bibinfo{author}{\bibfnamefont{Y.}~\bibnamefont{Samson}},
  \bibinfo{journal}{Science} \textbf{\bibinfo{volume}{284}},
  \bibinfo{pages}{2166} (\bibinfo{year}{1999}).

\bibitem[{\citenamefont{Abes et~al.}(2009)\citenamefont{Abes, Atkinson, Tanner,
  Charlton, Langridge, Hase, Ali, Marrows, Neudert, Hicken
  et~al.}}]{Abes:07C703}
\bibinfo{author}{\bibfnamefont{M.}~\bibnamefont{Abes}},
  \bibinfo{author}{\bibfnamefont{D.}~\bibnamefont{Atkinson}},
  \bibinfo{author}{\bibfnamefont{B.~K.} \bibnamefont{Tanner}},
  \bibinfo{author}{\bibfnamefont{T.}~\bibnamefont{Charlton}},
  \bibinfo{author}{\bibfnamefont{S.}~\bibnamefont{Langridge}},
  \bibinfo{author}{\bibfnamefont{T.~P.~A.} \bibnamefont{Hase}},
  \bibinfo{author}{\bibfnamefont{M.}~\bibnamefont{Ali}},
  \bibinfo{author}{\bibfnamefont{C.~H.} \bibnamefont{Marrows}},
  \bibinfo{author}{\bibfnamefont{A.}~\bibnamefont{Neudert}},
  \bibinfo{author}{\bibfnamefont{R.~J.} \bibnamefont{Hicken}},
  \bibnamefont{et~al.}, \bibinfo{journal}{J Appl Phys}
  \textbf{\bibinfo{volume}{105}}, \bibinfo{pages}{07C703}
  (\bibinfo{year}{2009}).

\bibitem[{\citenamefont{Tonnerre et~al.}(2008)\citenamefont{Tonnerre,
  De~Santis, Grenier, Tolentino, Langlais, Bontempi, Garc\'\i{}a-Fern\'andez,
  and Staub}}]{Tonnerre:157202}
\bibinfo{author}{\bibfnamefont{J.~M.} \bibnamefont{Tonnerre}},
  \bibinfo{author}{\bibfnamefont{M.}~\bibnamefont{De~Santis}},
  \bibinfo{author}{\bibfnamefont{S.}~\bibnamefont{Grenier}},
  \bibinfo{author}{\bibfnamefont{H.~C.~N.} \bibnamefont{Tolentino}},
  \bibinfo{author}{\bibfnamefont{V.}~\bibnamefont{Langlais}},
  \bibinfo{author}{\bibfnamefont{E.}~\bibnamefont{Bontempi}},
  \bibinfo{author}{\bibfnamefont{M.}~\bibnamefont{Garc\'\i{}a-Fern\'andez}},
  \bibnamefont{and} \bibinfo{author}{\bibfnamefont{U.}~\bibnamefont{Staub}},
  \bibinfo{journal}{Phys. Rev. Lett.} \textbf{\bibinfo{volume}{100}},
  \bibinfo{pages}{157202} (\bibinfo{year}{2008}).

\bibitem[{\citenamefont{Castleton and Altarelli}(2000)}]{Castleton:1033}
\bibinfo{author}{\bibfnamefont{C.~W.~M.} \bibnamefont{Castleton}}
  \bibnamefont{and}
  \bibinfo{author}{\bibfnamefont{M.}~\bibnamefont{Altarelli}},
  \bibinfo{journal}{Phys Rev B} \textbf{\bibinfo{volume}{62}},
  \bibinfo{pages}{1033} (\bibinfo{year}{2000}).

\bibitem[{\citenamefont{Wilkins
  et~al.}(2003{\natexlab{a}})\citenamefont{Wilkins, Hatton, Roper, Prabhakaran,
  and Boothroyd}}]{Wilkins:187201}
\bibinfo{author}{\bibfnamefont{S.~B.} \bibnamefont{Wilkins}},
  \bibinfo{author}{\bibfnamefont{P.~D.} \bibnamefont{Hatton}},
  \bibinfo{author}{\bibfnamefont{M.~D.} \bibnamefont{Roper}},
  \bibinfo{author}{\bibfnamefont{D.}~\bibnamefont{Prabhakaran}},
  \bibnamefont{and}
  \bibinfo{author}{\bibfnamefont{A.}~\bibnamefont{Boothroyd}},
  \bibinfo{journal}{Phys. Rev. Lett.} \textbf{\bibinfo{volume}{90}},
  \bibinfo{pages}{187201} (\bibinfo{year}{2003}{\natexlab{a}}).

\bibitem[{\citenamefont{Wilkins
  et~al.}(2003{\natexlab{b}})\citenamefont{Wilkins, Spencer, Hatton, Collins,
  Roper, Prabhakaran, and Boothroyd}}]{Wilkins:167205}
\bibinfo{author}{\bibfnamefont{S.~B.} \bibnamefont{Wilkins}},
  \bibinfo{author}{\bibfnamefont{P.~D.} \bibnamefont{Spencer}},
  \bibinfo{author}{\bibfnamefont{P.~D.} \bibnamefont{Hatton}},
  \bibinfo{author}{\bibfnamefont{S.~P.} \bibnamefont{Collins}},
  \bibinfo{author}{\bibfnamefont{M.~D.} \bibnamefont{Roper}},
  \bibinfo{author}{\bibfnamefont{D.}~\bibnamefont{Prabhakaran}},
  \bibnamefont{and} \bibinfo{author}{\bibfnamefont{A.~T.}
  \bibnamefont{Boothroyd}}, \bibinfo{journal}{Phys. Rev. Lett.}
  \textbf{\bibinfo{volume}{91}}, \bibinfo{pages}{167205}
  (\bibinfo{year}{2003}{\natexlab{b}}).

\bibitem[{\citenamefont{Dhesi et~al.}(2004)\citenamefont{Dhesi, Mirone, Nadai,
  Ohresser, Bencok, Brookes, Reutler, Revcolevschi, Tagliaferri, Toulemonde
  et~al.}}]{Dhesi:056403}
\bibinfo{author}{\bibfnamefont{S.~S.} \bibnamefont{Dhesi}},
  \bibinfo{author}{\bibfnamefont{A.}~\bibnamefont{Mirone}},
  \bibinfo{author}{\bibfnamefont{C.~D.} \bibnamefont{Nadai}},
  \bibinfo{author}{\bibfnamefont{P.}~\bibnamefont{Ohresser}},
  \bibinfo{author}{\bibfnamefont{P.}~\bibnamefont{Bencok}},
  \bibinfo{author}{\bibfnamefont{N.~B.} \bibnamefont{Brookes}},
  \bibinfo{author}{\bibfnamefont{P.}~\bibnamefont{Reutler}},
  \bibinfo{author}{\bibfnamefont{A.}~\bibnamefont{Revcolevschi}},
  \bibinfo{author}{\bibfnamefont{A.}~\bibnamefont{Tagliaferri}},
  \bibinfo{author}{\bibfnamefont{O.}~\bibnamefont{Toulemonde}},
  \bibnamefont{et~al.}, \bibinfo{journal}{Phys. Rev. Lett.}
  \textbf{\bibinfo{volume}{92}}, \bibinfo{pages}{056403}
  (\bibinfo{year}{2004}).

\bibitem[{\citenamefont{Thomas et~al.}(2004)\citenamefont{Thomas, Hill,
  Grenier, Kim, Abbamonte, Venema, Rusydi, Tomioka, Tokura, McMorrow
  et~al.}}]{Thomas:237204}
\bibinfo{author}{\bibfnamefont{K.~J.} \bibnamefont{Thomas}},
  \bibinfo{author}{\bibfnamefont{J.~P.} \bibnamefont{Hill}},
  \bibinfo{author}{\bibfnamefont{S.}~\bibnamefont{Grenier}},
  \bibinfo{author}{\bibfnamefont{J.~Y.} \bibnamefont{Kim}},
  \bibinfo{author}{\bibfnamefont{P.}~\bibnamefont{Abbamonte}},
  \bibinfo{author}{\bibfnamefont{L.}~\bibnamefont{Venema}},
  \bibinfo{author}{\bibfnamefont{A.}~\bibnamefont{Rusydi}},
  \bibinfo{author}{\bibfnamefont{Y.}~\bibnamefont{Tomioka}},
  \bibinfo{author}{\bibfnamefont{Y.}~\bibnamefont{Tokura}},
  \bibinfo{author}{\bibfnamefont{D.~F.} \bibnamefont{McMorrow}},
  \bibnamefont{et~al.}, \bibinfo{journal}{Phys. Rev. Lett.}
  \textbf{\bibinfo{volume}{92}}, \bibinfo{pages}{237204}
  (\bibinfo{year}{2004}).

\bibitem[{\citenamefont{Grenier et~al.}(2007)\citenamefont{Grenier, Thomas,
  Hill, Staub, Bodenthin, Garcia-Fernandez, Scagnoli, Kiryukhin, Cheong, Kim
  et~al.}}]{Grenier:206403}
\bibinfo{author}{\bibfnamefont{S.}~\bibnamefont{Grenier}},
  \bibinfo{author}{\bibfnamefont{K.~J.} \bibnamefont{Thomas}},
  \bibinfo{author}{\bibfnamefont{J.~P.} \bibnamefont{Hill}},
  \bibinfo{author}{\bibfnamefont{U.}~\bibnamefont{Staub}},
  \bibinfo{author}{\bibfnamefont{Y.}~\bibnamefont{Bodenthin}},
  \bibinfo{author}{\bibfnamefont{M.}~\bibnamefont{Garcia-Fernandez}},
  \bibinfo{author}{\bibfnamefont{V.}~\bibnamefont{Scagnoli}},
  \bibinfo{author}{\bibfnamefont{V.}~\bibnamefont{Kiryukhin}},
  \bibinfo{author}{\bibfnamefont{S.-W.} \bibnamefont{Cheong}},
  \bibinfo{author}{\bibfnamefont{B.~G.} \bibnamefont{Kim}},
  \bibnamefont{et~al.}, \bibinfo{journal}{Phys. Rev. Lett.}
  \textbf{\bibinfo{volume}{99}}, \bibinfo{pages}{206403}
  (\bibinfo{year}{2007}).

\bibitem[{\citenamefont{Wilkins et~al.}(2005)\citenamefont{Wilkins, Stojic,
  Beale, Binggeli, Castleton, Bencok, Prabhakaran, Boothroyd, Hatton, and
  Altarelli}}]{Wilkins:245102}
\bibinfo{author}{\bibfnamefont{S.~B.} \bibnamefont{Wilkins}},
  \bibinfo{author}{\bibfnamefont{N.}~\bibnamefont{Stojic}},
  \bibinfo{author}{\bibfnamefont{T.~A.~W.} \bibnamefont{Beale}},
  \bibinfo{author}{\bibfnamefont{N.}~\bibnamefont{Binggeli}},
  \bibinfo{author}{\bibfnamefont{C.}~\bibnamefont{Castleton}},
  \bibinfo{author}{\bibfnamefont{P.}~\bibnamefont{Bencok}},
  \bibinfo{author}{\bibfnamefont{D.}~\bibnamefont{Prabhakaran}},
  \bibinfo{author}{\bibfnamefont{A.~T.} \bibnamefont{Boothroyd}},
  \bibinfo{author}{\bibfnamefont{P.~D.} \bibnamefont{Hatton}},
  \bibnamefont{and}
  \bibinfo{author}{\bibfnamefont{M.}~\bibnamefont{Altarelli}},
  \bibinfo{journal}{Phys Rev B} \textbf{\bibinfo{volume}{71}},
  \bibinfo{pages}{245102} (\bibinfo{year}{2005}).

\bibitem[{\citenamefont{Staub et~al.}(2005)\citenamefont{Staub, Scagnoli,
  Mulders, Katsumata, Honda, Grimmer, Horisberger, and
  Tonnerre}}]{Staub:214421}
\bibinfo{author}{\bibfnamefont{U.}~\bibnamefont{Staub}},
  \bibinfo{author}{\bibfnamefont{V.}~\bibnamefont{Scagnoli}},
  \bibinfo{author}{\bibfnamefont{A.~M.} \bibnamefont{Mulders}},
  \bibinfo{author}{\bibfnamefont{K.}~\bibnamefont{Katsumata}},
  \bibinfo{author}{\bibfnamefont{Z.}~\bibnamefont{Honda}},
  \bibinfo{author}{\bibfnamefont{H.}~\bibnamefont{Grimmer}},
  \bibinfo{author}{\bibfnamefont{M.}~\bibnamefont{Horisberger}},
  \bibnamefont{and} \bibinfo{author}{\bibfnamefont{J.~M.}
  \bibnamefont{Tonnerre}}, \bibinfo{journal}{Phys Rev B}
  \textbf{\bibinfo{volume}{71}}, \bibinfo{pages}{214421}
  (\bibinfo{year}{2005}).

\bibitem[{\citenamefont{Wilkins et~al.}(2006)\citenamefont{Wilkins, Stojic,
  Beale, Binggeli, Hatton, Bencok, Stanescu, Mitchell, Abbamonte, and
  Altarelli}}]{Wilkins:L323}
\bibinfo{author}{\bibfnamefont{S.~B.} \bibnamefont{Wilkins}},
  \bibinfo{author}{\bibfnamefont{N.}~\bibnamefont{Stojic}},
  \bibinfo{author}{\bibfnamefont{T.~A.~W.} \bibnamefont{Beale}},
  \bibinfo{author}{\bibfnamefont{N.}~\bibnamefont{Binggeli}},
  \bibinfo{author}{\bibfnamefont{P.~D.} \bibnamefont{Hatton}},
  \bibinfo{author}{\bibfnamefont{P.}~\bibnamefont{Bencok}},
  \bibinfo{author}{\bibfnamefont{S.}~\bibnamefont{Stanescu}},
  \bibinfo{author}{\bibfnamefont{J.~F.} \bibnamefont{Mitchell}},
  \bibinfo{author}{\bibfnamefont{P.}~\bibnamefont{Abbamonte}},
  \bibnamefont{and}
  \bibinfo{author}{\bibfnamefont{M.}~\bibnamefont{Altarelli}},
  \bibinfo{journal}{J Phys C} \textbf{\bibinfo{volume}{18}},
  \bibinfo{pages}{L323} (\bibinfo{year}{2006}).

\bibitem[{\citenamefont{Herrero-Martin
  et~al.}(2006)\citenamefont{Herrero-Martin, Garcia, Subias, Blasco, Sanchez,
  and Stanescu}}]{HerreroMartin:224407}
\bibinfo{author}{\bibfnamefont{J.}~\bibnamefont{Herrero-Martin}},
  \bibinfo{author}{\bibfnamefont{J.}~\bibnamefont{Garcia}},
  \bibinfo{author}{\bibfnamefont{G.}~\bibnamefont{Subias}},
  \bibinfo{author}{\bibfnamefont{J.}~\bibnamefont{Blasco}},
  \bibinfo{author}{\bibfnamefont{M.~C.} \bibnamefont{Sanchez}},
  \bibnamefont{and} \bibinfo{author}{\bibfnamefont{S.}~\bibnamefont{Stanescu}},
  \bibinfo{journal}{Phys Rev B} \textbf{\bibinfo{volume}{73}},
  \bibinfo{pages}{224407} (\bibinfo{year}{2006}).

\bibitem[{\citenamefont{Beale et~al.}(2009)\citenamefont{Beale, Bland, Johnson,
  Hatton, Cezar, Dhesi, von~von Zimmermann, Prabhakaran, and
  Boothroyd}}]{Beale:054433}
\bibinfo{author}{\bibfnamefont{T.~A.~W.} \bibnamefont{Beale}},
  \bibinfo{author}{\bibfnamefont{S.~R.} \bibnamefont{Bland}},
  \bibinfo{author}{\bibfnamefont{R.~D.} \bibnamefont{Johnson}},
  \bibinfo{author}{\bibfnamefont{P.~D.} \bibnamefont{Hatton}},
  \bibinfo{author}{\bibfnamefont{J.~C.} \bibnamefont{Cezar}},
  \bibinfo{author}{\bibfnamefont{S.~S.} \bibnamefont{Dhesi}},
  \bibinfo{author}{\bibfnamefont{M.}~\bibnamefont{von~von Zimmermann}},
  \bibinfo{author}{\bibfnamefont{D.}~\bibnamefont{Prabhakaran}},
  \bibnamefont{and} \bibinfo{author}{\bibfnamefont{A.~T.}
  \bibnamefont{Boothroyd}}, \bibinfo{journal}{Phys Rev B}
  \textbf{\bibinfo{volume}{79}}, \bibinfo{pages}{054433}
  (\bibinfo{year}{2009}).

\bibitem[{\citenamefont{Stojic et~al.}(2005)\citenamefont{Stojic, Binggeli, and
  Altarelli}}]{Stojic:104108}
\bibinfo{author}{\bibfnamefont{N.}~\bibnamefont{Stojic}},
  \bibinfo{author}{\bibfnamefont{N.}~\bibnamefont{Binggeli}}, \bibnamefont{and}
  \bibinfo{author}{\bibfnamefont{M.}~\bibnamefont{Altarelli}},
  \bibinfo{journal}{Phys Rev B} \textbf{\bibinfo{volume}{72}},
  \bibinfo{pages}{104108} (\bibinfo{year}{2005}).

\bibitem[{\citenamefont{Mirone et~al.}(2006)\citenamefont{Mirone, Dhesi, and
  van~der Laan}}]{Mirone:23}
\bibinfo{author}{\bibfnamefont{A.}~\bibnamefont{Mirone}},
  \bibinfo{author}{\bibfnamefont{S.~S.} \bibnamefont{Dhesi}}, \bibnamefont{and}
  \bibinfo{author}{\bibfnamefont{G.}~\bibnamefont{van~der Laan}},
  \bibinfo{journal}{Eur. Phys. J. B} \textbf{\bibinfo{volume}{53}},
  \bibinfo{pages}{23} (\bibinfo{year}{2006}).

\bibitem[{\citenamefont{Abbamonte et~al.}(2005)\citenamefont{Abbamonte, Rusydi,
  Smadici, Gu, Sawatzky, and Feng}}]{Abbamonte:155}
\bibinfo{author}{\bibfnamefont{P.}~\bibnamefont{Abbamonte}},
  \bibinfo{author}{\bibfnamefont{A.}~\bibnamefont{Rusydi}},
  \bibinfo{author}{\bibfnamefont{S.}~\bibnamefont{Smadici}},
  \bibinfo{author}{\bibfnamefont{G.~D.} \bibnamefont{Gu}},
  \bibinfo{author}{\bibfnamefont{G.~A.} \bibnamefont{Sawatzky}},
  \bibnamefont{and} \bibinfo{author}{\bibfnamefont{D.~L.} \bibnamefont{Feng}},
  \bibinfo{journal}{Nat. Phys.} \textbf{\bibinfo{volume}{1}},
  \bibinfo{pages}{155} (\bibinfo{year}{2005}).

\bibitem[{\citenamefont{Abbamonte et~al.}(2002)\citenamefont{Abbamonte, Venema,
  Rusydi, Sawatzky, Logvenov, and Bozovic}}]{Abbamonte:581}
\bibinfo{author}{\bibfnamefont{P.}~\bibnamefont{Abbamonte}},
  \bibinfo{author}{\bibfnamefont{L.}~\bibnamefont{Venema}},
  \bibinfo{author}{\bibfnamefont{A.}~\bibnamefont{Rusydi}},
  \bibinfo{author}{\bibfnamefont{G.~A.} \bibnamefont{Sawatzky}},
  \bibinfo{author}{\bibfnamefont{G.}~\bibnamefont{Logvenov}}, \bibnamefont{and}
  \bibinfo{author}{\bibfnamefont{I.}~\bibnamefont{Bozovic}},
  \bibinfo{journal}{Science} \textbf{\bibinfo{volume}{297}},
  \bibinfo{pages}{581} (\bibinfo{year}{2002}).

\bibitem[{\citenamefont{Fink et~al.}(2009)\citenamefont{Fink, Schierle,
  Weschke, Geck, Hawthorn, Soltwisch, Wadati, Wu, Durr, Wizent
  et~al.}}]{Fink:100502}
\bibinfo{author}{\bibfnamefont{J.}~\bibnamefont{Fink}},
  \bibinfo{author}{\bibfnamefont{E.}~\bibnamefont{Schierle}},
  \bibinfo{author}{\bibfnamefont{E.}~\bibnamefont{Weschke}},
  \bibinfo{author}{\bibfnamefont{J.}~\bibnamefont{Geck}},
  \bibinfo{author}{\bibfnamefont{D.}~\bibnamefont{Hawthorn}},
  \bibinfo{author}{\bibfnamefont{V.}~\bibnamefont{Soltwisch}},
  \bibinfo{author}{\bibfnamefont{H.}~\bibnamefont{Wadati}},
  \bibinfo{author}{\bibfnamefont{H.-H.} \bibnamefont{Wu}},
  \bibinfo{author}{\bibfnamefont{H.~A.} \bibnamefont{Durr}},
  \bibinfo{author}{\bibfnamefont{N.}~\bibnamefont{Wizent}},
  \bibnamefont{et~al.}, \bibinfo{journal}{Phys Rev B}
  \textbf{\bibinfo{volume}{79}}, \bibinfo{pages}{100502}
  (\bibinfo{year}{2009}).

\bibitem[{\citenamefont{Schussler-Langeheine
  et~al.}(2005)\citenamefont{Schussler-Langeheine, Schlappa, Tanaka, Hu, Chang,
  Schierle, Benomar, Ott, Weschke, Kaindl et~al.}}]{SchusslerLangeheine:156402}
\bibinfo{author}{\bibfnamefont{C.}~\bibnamefont{Schussler-Langeheine}},
  \bibinfo{author}{\bibfnamefont{J.}~\bibnamefont{Schlappa}},
  \bibinfo{author}{\bibfnamefont{A.}~\bibnamefont{Tanaka}},
  \bibinfo{author}{\bibfnamefont{Z.}~\bibnamefont{Hu}},
  \bibinfo{author}{\bibfnamefont{C.}~\bibnamefont{Chang}},
  \bibinfo{author}{\bibfnamefont{E.}~\bibnamefont{Schierle}},
  \bibinfo{author}{\bibfnamefont{M.}~\bibnamefont{Benomar}},
  \bibinfo{author}{\bibfnamefont{H.}~\bibnamefont{Ott}},
  \bibinfo{author}{\bibfnamefont{E.}~\bibnamefont{Weschke}},
  \bibinfo{author}{\bibfnamefont{G.}~\bibnamefont{Kaindl}},
  \bibnamefont{et~al.}, \bibinfo{journal}{Phys. Rev. Lett.}
  \textbf{\bibinfo{volume}{95}}, \bibinfo{pages}{156402}
  (\bibinfo{year}{2005}).

\bibitem[{\citenamefont{Scagnoli et~al.}(2006)\citenamefont{Scagnoli, Staub,
  Mulders, Janousch, Meijer, Hammerl, Tonnerre, and Stojic}}]{Scagnoli:100409}
\bibinfo{author}{\bibfnamefont{V.}~\bibnamefont{Scagnoli}},
  \bibinfo{author}{\bibfnamefont{U.}~\bibnamefont{Staub}},
  \bibinfo{author}{\bibfnamefont{A.~M.} \bibnamefont{Mulders}},
  \bibinfo{author}{\bibfnamefont{M.}~\bibnamefont{Janousch}},
  \bibinfo{author}{\bibfnamefont{G.~I.} \bibnamefont{Meijer}},
  \bibinfo{author}{\bibfnamefont{G.}~\bibnamefont{Hammerl}},
  \bibinfo{author}{\bibfnamefont{J.~M.} \bibnamefont{Tonnerre}},
  \bibnamefont{and} \bibinfo{author}{\bibfnamefont{N.}~\bibnamefont{Stojic}},
  \bibinfo{journal}{Phys Rev B} \textbf{\bibinfo{volume}{73}},
  \bibinfo{pages}{100409} (\bibinfo{year}{2006}).

\bibitem[{\citenamefont{Chang et~al.}(2009)\citenamefont{Chang, Hu, Wu, Burnus,
  Hollmann, Benomar, Lorenz, Tanaka, Lin, Hsieh et~al.}}]{Chang:116401}
\bibinfo{author}{\bibfnamefont{C.}~\bibnamefont{Chang}},
  \bibinfo{author}{\bibfnamefont{Z.}~\bibnamefont{Hu}},
  \bibinfo{author}{\bibfnamefont{H.}~\bibnamefont{Wu}},
  \bibinfo{author}{\bibfnamefont{T.}~\bibnamefont{Burnus}},
  \bibinfo{author}{\bibfnamefont{N.}~\bibnamefont{Hollmann}},
  \bibinfo{author}{\bibfnamefont{M.}~\bibnamefont{Benomar}},
  \bibinfo{author}{\bibfnamefont{T.}~\bibnamefont{Lorenz}},
  \bibinfo{author}{\bibfnamefont{A.}~\bibnamefont{Tanaka}},
  \bibinfo{author}{\bibfnamefont{H.-J.} \bibnamefont{Lin}},
  \bibinfo{author}{\bibfnamefont{H.~H.} \bibnamefont{Hsieh}},
  \bibnamefont{et~al.}, \bibinfo{journal}{Phys. Rev. Lett.}
  \textbf{\bibinfo{volume}{102}}, \bibinfo{pages}{116401}
  (\bibinfo{year}{2009}).

\bibitem[{\citenamefont{Beale et~al.}(2007)\citenamefont{Beale, Wilkins,
  Hatton, Abbamonte, Stanescu, and Paix{\~a}o}}]{Beale:174432}
\bibinfo{author}{\bibfnamefont{T.~A.~W.} \bibnamefont{Beale}},
  \bibinfo{author}{\bibfnamefont{S.~B.} \bibnamefont{Wilkins}},
  \bibinfo{author}{\bibfnamefont{P.~D.} \bibnamefont{Hatton}},
  \bibinfo{author}{\bibfnamefont{P.}~\bibnamefont{Abbamonte}},
  \bibinfo{author}{\bibfnamefont{S.}~\bibnamefont{Stanescu}}, \bibnamefont{and}
  \bibinfo{author}{\bibfnamefont{J.~A.} \bibnamefont{Paix{\~a}o}},
  \bibinfo{journal}{Phys Rev B} \textbf{\bibinfo{volume}{75}},
  \bibinfo{pages}{174432} (\bibinfo{year}{2007}).

\bibitem[{\citenamefont{Mulders et~al.}(2006)\citenamefont{Mulders, Staub,
  Scagnoli, Lovesey, Balcar, Nakamura, Kikkawa, van~der Laan, and
  Tonnerre}}]{Mulders:11195}
\bibinfo{author}{\bibfnamefont{A.~M.} \bibnamefont{Mulders}},
  \bibinfo{author}{\bibfnamefont{U.}~\bibnamefont{Staub}},
  \bibinfo{author}{\bibfnamefont{V.}~\bibnamefont{Scagnoli}},
  \bibinfo{author}{\bibfnamefont{S.~W.} \bibnamefont{Lovesey}},
  \bibinfo{author}{\bibfnamefont{E.}~\bibnamefont{Balcar}},
  \bibinfo{author}{\bibfnamefont{T.}~\bibnamefont{Nakamura}},
  \bibinfo{author}{\bibfnamefont{A.}~\bibnamefont{Kikkawa}},
  \bibinfo{author}{\bibfnamefont{G.}~\bibnamefont{van~der Laan}},
  \bibnamefont{and} \bibinfo{author}{\bibfnamefont{J.~M.}
  \bibnamefont{Tonnerre}}, \bibinfo{journal}{J Phys C}
  \textbf{\bibinfo{volume}{18}}, \bibinfo{pages}{11195} (\bibinfo{year}{2006}).

\bibitem[{\citenamefont{Mulders et~al.}(2007)\citenamefont{Mulders, Staub,
  Scagnoli, Tanaka, Kikkawa, Katsumata, and Tonnerre}}]{Mulders:184438}
\bibinfo{author}{\bibfnamefont{A.~M.} \bibnamefont{Mulders}},
  \bibinfo{author}{\bibfnamefont{U.}~\bibnamefont{Staub}},
  \bibinfo{author}{\bibfnamefont{V.}~\bibnamefont{Scagnoli}},
  \bibinfo{author}{\bibfnamefont{Y.}~\bibnamefont{Tanaka}},
  \bibinfo{author}{\bibfnamefont{A.}~\bibnamefont{Kikkawa}},
  \bibinfo{author}{\bibfnamefont{K.}~\bibnamefont{Katsumata}},
  \bibnamefont{and} \bibinfo{author}{\bibfnamefont{J.~M.}
  \bibnamefont{Tonnerre}}, \bibinfo{journal}{Phys Rev B}
  \textbf{\bibinfo{volume}{75}}, \bibinfo{pages}{184438}
  (\bibinfo{year}{2007}).

\bibitem[{\citenamefont{Koo et~al.}(2007)\citenamefont{Koo, Song, Ji, Lee,
  Park, Jang, Yang, Park, Jeong, Lee et~al.}}]{Koo:197601}
\bibinfo{author}{\bibfnamefont{J.}~\bibnamefont{Koo}},
  \bibinfo{author}{\bibfnamefont{C.}~\bibnamefont{Song}},
  \bibinfo{author}{\bibfnamefont{S.}~\bibnamefont{Ji}},
  \bibinfo{author}{\bibfnamefont{J.-S.} \bibnamefont{Lee}},
  \bibinfo{author}{\bibfnamefont{J.}~\bibnamefont{Park}},
  \bibinfo{author}{\bibfnamefont{T.-H.} \bibnamefont{Jang}},
  \bibinfo{author}{\bibfnamefont{C.-H.} \bibnamefont{Yang}},
  \bibinfo{author}{\bibfnamefont{J.-H.} \bibnamefont{Park}},
  \bibinfo{author}{\bibfnamefont{Y.~H.} \bibnamefont{Jeong}},
  \bibinfo{author}{\bibfnamefont{K.-B.} \bibnamefont{Lee}},
  \bibnamefont{et~al.}, \bibinfo{journal}{Phys. Rev. Lett.}
  \textbf{\bibinfo{volume}{99}}, \bibinfo{pages}{197601}
  (\bibinfo{year}{2007}).

\bibitem[{\citenamefont{Bodenthin et~al.}(2008)\citenamefont{Bodenthin, Staub,
  Garcia-Fernandez, Janoschek, Schlappa, Golovenchits, Sanina, and
  Lushnikov}}]{Bodenthin:027201}
\bibinfo{author}{\bibfnamefont{Y.}~\bibnamefont{Bodenthin}},
  \bibinfo{author}{\bibfnamefont{U.}~\bibnamefont{Staub}},
  \bibinfo{author}{\bibfnamefont{M.}~\bibnamefont{Garcia-Fernandez}},
  \bibinfo{author}{\bibfnamefont{M.}~\bibnamefont{Janoschek}},
  \bibinfo{author}{\bibfnamefont{J.}~\bibnamefont{Schlappa}},
  \bibinfo{author}{\bibfnamefont{E.~I.} \bibnamefont{Golovenchits}},
  \bibinfo{author}{\bibfnamefont{V.~A.} \bibnamefont{Sanina}},
  \bibnamefont{and} \bibinfo{author}{\bibfnamefont{S.~G.}
  \bibnamefont{Lushnikov}}, \bibinfo{journal}{Phys. Rev. Lett.}
  \textbf{\bibinfo{volume}{100}}, \bibinfo{pages}{027201}
  (\bibinfo{year}{2008}).

\bibitem[{\citenamefont{Okamoto et~al.}(2007)\citenamefont{Okamoto, Huang, Mou,
  Chao, Lin, Park, Cheong, and Chen}}]{Okamoto:157202}
\bibinfo{author}{\bibfnamefont{J.}~\bibnamefont{Okamoto}},
  \bibinfo{author}{\bibfnamefont{D.~J.} \bibnamefont{Huang}},
  \bibinfo{author}{\bibfnamefont{C.-Y.} \bibnamefont{Mou}},
  \bibinfo{author}{\bibfnamefont{K.~S.} \bibnamefont{Chao}},
  \bibinfo{author}{\bibfnamefont{H.-J.} \bibnamefont{Lin}},
  \bibinfo{author}{\bibfnamefont{S.}~\bibnamefont{Park}},
  \bibinfo{author}{\bibfnamefont{S.-W.} \bibnamefont{Cheong}},
  \bibnamefont{and} \bibinfo{author}{\bibfnamefont{C.~T.} \bibnamefont{Chen}},
  \bibinfo{journal}{Phys. Rev. Lett.} \textbf{\bibinfo{volume}{98}},
  \bibinfo{pages}{157202} (\bibinfo{year}{2007}).

\bibitem[{\citenamefont{Forrest et~al.}(2008)\citenamefont{Forrest, Bland,
  Wilkins, Walker, Beale, Hatton, Prabhakaran, Boothroyd, Mannix, Yakhou
  et~al.}}]{Forrest:422205}
\bibinfo{author}{\bibfnamefont{T.}~\bibnamefont{Forrest}},
  \bibinfo{author}{\bibfnamefont{S.~R.} \bibnamefont{Bland}},
  \bibinfo{author}{\bibfnamefont{S.~B.} \bibnamefont{Wilkins}},
  \bibinfo{author}{\bibfnamefont{H.~C.} \bibnamefont{Walker}},
  \bibinfo{author}{\bibfnamefont{T.~A.~W.} \bibnamefont{Beale}},
  \bibinfo{author}{\bibfnamefont{P.~D.} \bibnamefont{Hatton}},
  \bibinfo{author}{\bibfnamefont{D.}~\bibnamefont{Prabhakaran}},
  \bibinfo{author}{\bibfnamefont{A.}~\bibnamefont{Boothroyd}},
  \bibinfo{author}{\bibfnamefont{D.}~\bibnamefont{Mannix}},
  \bibinfo{author}{\bibfnamefont{F.}~\bibnamefont{Yakhou}},
  \bibnamefont{et~al.}, \bibinfo{journal}{J Phys C}
  \textbf{\bibinfo{volume}{20}}, \bibinfo{pages}{422205}
  (\bibinfo{year}{2008}).

\bibitem[{\citenamefont{Wilkins et~al.}(2009)\citenamefont{Wilkins, Forrest,
  Beale, Bland, Walker, Mannix, Yakhou, Prabhakaran, Boothroyd, Hill
  et~al.}}]{Wilkins:207602}
\bibinfo{author}{\bibfnamefont{S.~B.} \bibnamefont{Wilkins}},
  \bibinfo{author}{\bibfnamefont{T.~R.} \bibnamefont{Forrest}},
  \bibinfo{author}{\bibfnamefont{T.~A.~W.} \bibnamefont{Beale}},
  \bibinfo{author}{\bibfnamefont{S.~R.} \bibnamefont{Bland}},
  \bibinfo{author}{\bibfnamefont{H.~C.} \bibnamefont{Walker}},
  \bibinfo{author}{\bibfnamefont{D.}~\bibnamefont{Mannix}},
  \bibinfo{author}{\bibfnamefont{F.}~\bibnamefont{Yakhou}},
  \bibinfo{author}{\bibfnamefont{D.}~\bibnamefont{Prabhakaran}},
  \bibinfo{author}{\bibfnamefont{A.~T.} \bibnamefont{Boothroyd}},
  \bibinfo{author}{\bibfnamefont{J.~P.} \bibnamefont{Hill}},
  \bibnamefont{et~al.}, \bibinfo{journal}{Phys. Rev. Lett.}
  \textbf{\bibinfo{volume}{103}}, \bibinfo{pages}{207602}
  (\bibinfo{year}{2009}).

\bibitem[{\citenamefont{Ohtomo and Hwang}(2004)}]{Ohtomo:423}
\bibinfo{author}{\bibfnamefont{A.}~\bibnamefont{Ohtomo}} \bibnamefont{and}
  \bibinfo{author}{\bibfnamefont{H.}~\bibnamefont{Hwang}},
  \bibinfo{journal}{Nature} \textbf{\bibinfo{volume}{427}},
  \bibinfo{pages}{423} (\bibinfo{year}{2004}).

\bibitem[{\citenamefont{Chakhalian et~al.}(2007)\citenamefont{Chakhalian,
  Freeland, Habermeier, Cristiani, Khaliullin, van Veenendaal, and
  Keimer}}]{Chakhalian:1114}
\bibinfo{author}{\bibfnamefont{J.}~\bibnamefont{Chakhalian}},
  \bibinfo{author}{\bibfnamefont{J.~W.} \bibnamefont{Freeland}},
  \bibinfo{author}{\bibfnamefont{H.-U.} \bibnamefont{Habermeier}},
  \bibinfo{author}{\bibfnamefont{G.}~\bibnamefont{Cristiani}},
  \bibinfo{author}{\bibfnamefont{G.}~\bibnamefont{Khaliullin}},
  \bibinfo{author}{\bibfnamefont{M.}~\bibnamefont{van Veenendaal}},
  \bibnamefont{and} \bibinfo{author}{\bibfnamefont{B.}~\bibnamefont{Keimer}},
  \bibinfo{journal}{Science} \textbf{\bibinfo{volume}{318}},
  \bibinfo{pages}{1114} (\bibinfo{year}{2007}).

\bibitem[{\citenamefont{Chesnel et~al.}(2002)\citenamefont{Chesnel,
  Belakhovsky, Livet, Collins, van~der Laan, Dhesi, Attan\'{e}, and
  Marty}}]{chesnel02}
\bibinfo{author}{\bibfnamefont{K.}~\bibnamefont{Chesnel}},
  \bibinfo{author}{\bibfnamefont{M.}~\bibnamefont{Belakhovsky}},
  \bibinfo{author}{\bibfnamefont{F.}~\bibnamefont{Livet}},
  \bibinfo{author}{\bibfnamefont{S.~P.} \bibnamefont{Collins}},
  \bibinfo{author}{\bibfnamefont{G.}~\bibnamefont{van~der Laan}},
  \bibinfo{author}{\bibfnamefont{S.~S.} \bibnamefont{Dhesi}},
  \bibinfo{author}{\bibfnamefont{J.~P.} \bibnamefont{Attan\'{e}}},
  \bibnamefont{and} \bibinfo{author}{\bibfnamefont{A.}~\bibnamefont{Marty}},
  \bibinfo{journal}{Phys. Rev. B} \textbf{\bibinfo{volume}{66}},
  \bibinfo{pages}{172404} (\bibinfo{year}{2002}).

\bibitem[{\citenamefont{Chesnel et~al.}(2004)\citenamefont{Chesnel,
  Belakhovsky, van~der Laan, Livet, Marty, Beutier, Collins, and
  Haznar}}]{chesnel04}
\bibinfo{author}{\bibfnamefont{K.}~\bibnamefont{Chesnel}},
  \bibinfo{author}{\bibfnamefont{M.}~\bibnamefont{Belakhovsky}},
  \bibinfo{author}{\bibfnamefont{G.}~\bibnamefont{van~der Laan}},
  \bibinfo{author}{\bibfnamefont{F.}~\bibnamefont{Livet}},
  \bibinfo{author}{\bibfnamefont{A.}~\bibnamefont{Marty}},
  \bibinfo{author}{\bibfnamefont{G.}~\bibnamefont{Beutier}},
  \bibinfo{author}{\bibfnamefont{S.~P.} \bibnamefont{Collins}},
  \bibnamefont{and} \bibinfo{author}{\bibfnamefont{A.}~\bibnamefont{Haznar}},
  \bibinfo{journal}{Phys. Rev. B} \textbf{\bibinfo{volume}{70}},
  \bibinfo{pages}{180402} (\bibinfo{year}{2004}).

\bibitem[{\citenamefont{Dudzik et~al.}(2000)\citenamefont{Dudzik, Dhesi, DŸrr,
  Collins, Roper, van~der Laan, Chesnel, Belakhovsky, Marty, and
  Samson}}]{dudzik}
\bibinfo{author}{\bibfnamefont{E.}~\bibnamefont{Dudzik}},
  \bibinfo{author}{\bibfnamefont{S.~S.} \bibnamefont{Dhesi}},
  \bibinfo{author}{\bibfnamefont{H.~A.} \bibnamefont{DŸrr}},
  \bibinfo{author}{\bibfnamefont{S.~P.} \bibnamefont{Collins}},
  \bibinfo{author}{\bibfnamefont{M.~D.} \bibnamefont{Roper}},
  \bibinfo{author}{\bibfnamefont{G.}~\bibnamefont{van~der Laan}},
  \bibinfo{author}{\bibfnamefont{K.}~\bibnamefont{Chesnel}},
  \bibinfo{author}{\bibfnamefont{M.}~\bibnamefont{Belakhovsky}},
  \bibinfo{author}{\bibfnamefont{A.}~\bibnamefont{Marty}}, \bibnamefont{and}
  \bibinfo{author}{\bibfnamefont{Y.}~\bibnamefont{Samson}},
  \bibinfo{journal}{Phys. Rev. B} \textbf{\bibinfo{volume}{62}},
  \bibinfo{pages}{5779} (\bibinfo{year}{2000}).

\bibitem[{\citenamefont{Haznar et~al.}(2004)\citenamefont{Haznar, van~der Laan,
  Collins, Vaz, Bland, and Dhesi}}]{haznar}
\bibinfo{author}{\bibfnamefont{A.}~\bibnamefont{Haznar}},
  \bibinfo{author}{\bibfnamefont{G.}~\bibnamefont{van~der Laan}},
  \bibinfo{author}{\bibfnamefont{S.~P.} \bibnamefont{Collins}},
  \bibinfo{author}{\bibfnamefont{C.~A.~F.} \bibnamefont{Vaz}},
  \bibinfo{author}{\bibfnamefont{J.~A.~C.} \bibnamefont{Bland}},
  \bibnamefont{and} \bibinfo{author}{\bibfnamefont{S.~S.} \bibnamefont{Dhesi}},
  \bibinfo{journal}{J. Synch. Rad.} \textbf{\bibinfo{volume}{11}},
  \bibinfo{pages}{254} (\bibinfo{year}{2004}).

\bibitem[{\citenamefont{Ogrin et~al.}(2008)\citenamefont{Ogrin, Sirotkin,
  van~der Laan, Beutier, Ross, Jung, and Menon}}]{ogrin}
\bibinfo{author}{\bibfnamefont{F.~Y.} \bibnamefont{Ogrin}},
  \bibinfo{author}{\bibfnamefont{E.}~\bibnamefont{Sirotkin}},
  \bibinfo{author}{\bibfnamefont{G.}~\bibnamefont{van~der Laan}},
  \bibinfo{author}{\bibfnamefont{G.}~\bibnamefont{Beutier}},
  \bibinfo{author}{\bibfnamefont{C.~A.} \bibnamefont{Ross}},
  \bibinfo{author}{\bibfnamefont{W.}~\bibnamefont{Jung}}, \bibnamefont{and}
  \bibinfo{author}{\bibfnamefont{R.}~\bibnamefont{Menon}}, \bibinfo{journal}{J.
  Appl. Phys.} \textbf{\bibinfo{volume}{103}}, \bibinfo{pages}{07E909}
  (\bibinfo{year}{2008}).

\bibitem[{\citenamefont{Roper et~al.}(2001)\citenamefont{Roper, van~der Laan,
  Dürr, Dudzik, Collins, Miller, and Thompson}}]{Roper:1101}
\bibinfo{author}{\bibfnamefont{M.~D.} \bibnamefont{Roper}},
  \bibinfo{author}{\bibfnamefont{G.}~\bibnamefont{van~der Laan}},
  \bibinfo{author}{\bibfnamefont{H.~A.} \bibnamefont{Dürr}},
  \bibinfo{author}{\bibfnamefont{E.}~\bibnamefont{Dudzik}},
  \bibinfo{author}{\bibfnamefont{S.~P.} \bibnamefont{Collins}},
  \bibinfo{author}{\bibfnamefont{M.~C.} \bibnamefont{Miller}},
  \bibnamefont{and} \bibinfo{author}{\bibfnamefont{S.~P.}
  \bibnamefont{Thompson}}, \bibinfo{journal}{Nuclear Instruments and Methods in
  Physics Research Section A: Accelerators, Spectrometers, Detectors and
  Associated Equipment} \textbf{\bibinfo{volume}{467-468}},
  \bibinfo{pages}{1101} (\bibinfo{year}{2001}).

\bibitem[{\citenamefont{Staub et~al.}(2008)\citenamefont{Staub, Scagnoli,
  Bodenthin, Garcia-Fernandez, Wetter, Mulders, Grimmer, and
  Horisberger}}]{Staub:469}
\bibinfo{author}{\bibfnamefont{U.}~\bibnamefont{Staub}},
  \bibinfo{author}{\bibfnamefont{V.}~\bibnamefont{Scagnoli}},
  \bibinfo{author}{\bibfnamefont{Y.}~\bibnamefont{Bodenthin}},
  \bibinfo{author}{\bibfnamefont{M.}~\bibnamefont{Garcia-Fernandez}},
  \bibinfo{author}{\bibfnamefont{R.}~\bibnamefont{Wetter}},
  \bibinfo{author}{\bibfnamefont{A.~M.} \bibnamefont{Mulders}},
  \bibinfo{author}{\bibfnamefont{H.}~\bibnamefont{Grimmer}}, \bibnamefont{and}
  \bibinfo{author}{\bibfnamefont{M.}~\bibnamefont{Horisberger}},
  \bibinfo{journal}{Journal of Synchrotron Radiation}
  \textbf{\bibinfo{volume}{15}}, \bibinfo{pages}{469} (\bibinfo{year}{2008}).

\bibitem[{\citenamefont{Beutier et~al.}(2007)\citenamefont{Beutier, Marty,
  Livet, van~der Laan, Stanescu, and Bencok}}]{Beutier:093901}
\bibinfo{author}{\bibfnamefont{G.}~\bibnamefont{Beutier}},
  \bibinfo{author}{\bibfnamefont{A.}~\bibnamefont{Marty}},
  \bibinfo{author}{\bibfnamefont{F.}~\bibnamefont{Livet}},
  \bibinfo{author}{\bibfnamefont{G.}~\bibnamefont{van~der Laan}},
  \bibinfo{author}{\bibfnamefont{S.}~\bibnamefont{Stanescu}}, \bibnamefont{and}
  \bibinfo{author}{\bibfnamefont{P.}~\bibnamefont{Bencok}},
  \bibinfo{journal}{Rev. Sci. Instrum.} \textbf{\bibinfo{volume}{78}},
  \bibinfo{pages}{093901} (\bibinfo{year}{2007}).

\bibitem[{\citenamefont{Takeuchi et~al.}(2009)\citenamefont{Takeuchi, Chainani,
  Takata, Tanaka, Oura, Tsubota, Senba, Ohashi, Mochiku, Hirata
  et~al.}}]{Takeuchi:023905}
\bibinfo{author}{\bibfnamefont{T.}~\bibnamefont{Takeuchi}},
  \bibinfo{author}{\bibfnamefont{A.}~\bibnamefont{Chainani}},
  \bibinfo{author}{\bibfnamefont{Y.}~\bibnamefont{Takata}},
  \bibinfo{author}{\bibfnamefont{Y.}~\bibnamefont{Tanaka}},
  \bibinfo{author}{\bibfnamefont{M.}~\bibnamefont{Oura}},
  \bibinfo{author}{\bibfnamefont{M.}~\bibnamefont{Tsubota}},
  \bibinfo{author}{\bibfnamefont{Y.}~\bibnamefont{Senba}},
  \bibinfo{author}{\bibfnamefont{H.}~\bibnamefont{Ohashi}},
  \bibinfo{author}{\bibfnamefont{T.}~\bibnamefont{Mochiku}},
  \bibinfo{author}{\bibfnamefont{K.}~\bibnamefont{Hirata}},
  \bibnamefont{et~al.}, \bibinfo{journal}{Rev. Sci. Instrum.}
  \textbf{\bibinfo{volume}{80}}, \bibinfo{pages}{023905}
  (\bibinfo{year}{2009}).

\bibitem[{\citenamefont{Grabis et~al.}(2003)\citenamefont{Grabis, Nefedov, and
  Zabel}}]{Grabis:4048}
\bibinfo{author}{\bibfnamefont{J.}~\bibnamefont{Grabis}},
  \bibinfo{author}{\bibfnamefont{A.}~\bibnamefont{Nefedov}}, \bibnamefont{and}
  \bibinfo{author}{\bibfnamefont{H.}~\bibnamefont{Zabel}},
  \bibinfo{journal}{Rev. Sci. Instrum.} \textbf{\bibinfo{volume}{74}},
  \bibinfo{pages}{4048} (\bibinfo{year}{2003}).

\bibitem[{Toy()}]{Toyama}
\bibinfo{howpublished}{\url{http://www.toyama-jp.com}}.

\bibitem[{\citenamefont{Murakami et~al.}(2010)\citenamefont{Murakami, Konno,
  Arima, Shindo, and Suzuki}}]{Hill:140102}
\bibinfo{author}{\bibfnamefont{Y.}~\bibnamefont{Murakami}},
  \bibinfo{author}{\bibfnamefont{S.}~\bibnamefont{Konno}},
  \bibinfo{author}{\bibfnamefont{T.}~\bibnamefont{Arima}},
  \bibinfo{author}{\bibfnamefont{D.}~\bibnamefont{Shindo}}, \bibnamefont{and}
  \bibinfo{author}{\bibfnamefont{T.}~\bibnamefont{Suzuki}},
  \bibinfo{journal}{Phys. Rev. B} \textbf{\bibinfo{volume}{81}},
  \bibinfo{pages}{140102} (\bibinfo{year}{2010}).

\bibitem[{\citenamefont{Scagnoli et~al.}(2009)\citenamefont{Scagnoli, Mazzoli,
  Detelfs, Bernard, Fondacaro, Paolasini, Fabrizi, and
  de~Bergevin}}]{Scagnoli:778}
\bibinfo{author}{\bibfnamefont{V.}~\bibnamefont{Scagnoli}},
  \bibinfo{author}{\bibfnamefont{C.}~\bibnamefont{Mazzoli}},
  \bibinfo{author}{\bibfnamefont{C.}~\bibnamefont{Detelfs}},
  \bibinfo{author}{\bibfnamefont{P.}~\bibnamefont{Bernard}},
  \bibinfo{author}{\bibfnamefont{A.}~\bibnamefont{Fondacaro}},
  \bibinfo{author}{\bibfnamefont{L.}~\bibnamefont{Paolasini}},
  \bibinfo{author}{\bibfnamefont{F.}~\bibnamefont{Fabrizi}}, \bibnamefont{and}
  \bibinfo{author}{\bibfnamefont{F.}~\bibnamefont{de~Bergevin}},
  \bibinfo{journal}{J. Sync. Rad.} \textbf{\bibinfo{volume}{16}},
  \bibinfo{pages}{778} (\bibinfo{year}{2009}).

\bibitem[{\citenamefont{Mazzoli et~al.}(2007)\citenamefont{Mazzoli, Wilkins,
  Di~Matteo, Detlefs, Detlefs, Scagnoli, Paolasini, and
  Ghigna}}]{Mazzoli:195118}
\bibinfo{author}{\bibfnamefont{C.}~\bibnamefont{Mazzoli}},
  \bibinfo{author}{\bibfnamefont{S.~B.} \bibnamefont{Wilkins}},
  \bibinfo{author}{\bibfnamefont{S.}~\bibnamefont{Di~Matteo}},
  \bibinfo{author}{\bibfnamefont{B.}~\bibnamefont{Detlefs}},
  \bibinfo{author}{\bibfnamefont{C.}~\bibnamefont{Detlefs}},
  \bibinfo{author}{\bibfnamefont{V.}~\bibnamefont{Scagnoli}},
  \bibinfo{author}{\bibfnamefont{L.}~\bibnamefont{Paolasini}},
  \bibnamefont{and} \bibinfo{author}{\bibfnamefont{P.}~\bibnamefont{Ghigna}},
  \bibinfo{journal}{Phys. Rev. B} \textbf{\bibinfo{volume}{76}},
  \bibinfo{pages}{195118} (\bibinfo{year}{2007}).

\bibitem[{\citenamefont{Johnson et~al.}(2008)\citenamefont{Johnson, Bland,
  Mazzoli, Beale, Du, Detlefs, Wilkins, and Hatton}}]{Johnson:104407}
\bibinfo{author}{\bibfnamefont{R.~D.} \bibnamefont{Johnson}},
  \bibinfo{author}{\bibfnamefont{S.~R.} \bibnamefont{Bland}},
  \bibinfo{author}{\bibfnamefont{C.}~\bibnamefont{Mazzoli}},
  \bibinfo{author}{\bibfnamefont{T.~A.~W.} \bibnamefont{Beale}},
  \bibinfo{author}{\bibfnamefont{C.-H.} \bibnamefont{Du}},
  \bibinfo{author}{\bibfnamefont{C.}~\bibnamefont{Detlefs}},
  \bibinfo{author}{\bibfnamefont{S.~B.} \bibnamefont{Wilkins}},
  \bibnamefont{and} \bibinfo{author}{\bibfnamefont{P.~D.}
  \bibnamefont{Hatton}}, \bibinfo{journal}{Phys. Rev. B}
  \textbf{\bibinfo{volume}{78}}, \bibinfo{pages}{104407}
  (\bibinfo{year}{2008}).

\bibitem[{epi()}]{epics}
\bibinfo{howpublished}{\url{http://www.aps.anl.gov/epics}}.

\bibitem[{gda()}]{gda}
\bibinfo{howpublished}{\url{http://www.opengda.org}}.

\bibitem[{\citenamefont{Bj\"{o}rck and Andersson}(2007)}]{bjorck:1174}
\bibinfo{author}{\bibfnamefont{M.}~\bibnamefont{Bj\"{o}rck}} \bibnamefont{and}
  \bibinfo{author}{\bibfnamefont{G.}~\bibnamefont{Andersson}},
  \bibinfo{journal}{J. Appl. Cryst.} \textbf{\bibinfo{volume}{40}},
  \bibinfo{pages}{1174} (\bibinfo{year}{2007}).

\bibitem[{\citenamefont{Parratt}(1954)}]{Parratt:359}
\bibinfo{author}{\bibfnamefont{L.~G.} \bibnamefont{Parratt}},
  \bibinfo{journal}{Phys. Rev.} \textbf{\bibinfo{volume}{95}},
  \bibinfo{pages}{359} (\bibinfo{year}{1954}).

\bibitem[{\citenamefont{Sch\"utz et~al.}(1987)\citenamefont{Sch\"utz, Wagner,
  Wilhelm, Kienle, Zeller, Frahm, and Materlik}}]{Schutz:737}
\bibinfo{author}{\bibfnamefont{G.}~\bibnamefont{Sch\"utz}},
  \bibinfo{author}{\bibfnamefont{W.}~\bibnamefont{Wagner}},
  \bibinfo{author}{\bibfnamefont{W.}~\bibnamefont{Wilhelm}},
  \bibinfo{author}{\bibfnamefont{P.}~\bibnamefont{Kienle}},
  \bibinfo{author}{\bibfnamefont{R.}~\bibnamefont{Zeller}},
  \bibinfo{author}{\bibfnamefont{R.}~\bibnamefont{Frahm}}, \bibnamefont{and}
  \bibinfo{author}{\bibfnamefont{G.}~\bibnamefont{Materlik}},
  \bibinfo{journal}{Phys. Rev. Lett.} \textbf{\bibinfo{volume}{58}},
  \bibinfo{pages}{737} (\bibinfo{year}{1987}).

\bibitem[{\citenamefont{Mulders et~al.}(2009)\citenamefont{Mulders, Lawrence,
  Staub, Garcia-Fernandez, Scagnoli, Mazzoli, Pomjakushina, Conder, and
  Wang}}]{Mulders:077602}
\bibinfo{author}{\bibfnamefont{A.~M.} \bibnamefont{Mulders}},
  \bibinfo{author}{\bibfnamefont{S.~M.} \bibnamefont{Lawrence}},
  \bibinfo{author}{\bibfnamefont{U.}~\bibnamefont{Staub}},
  \bibinfo{author}{\bibfnamefont{M.}~\bibnamefont{Garcia-Fernandez}},
  \bibinfo{author}{\bibfnamefont{V.}~\bibnamefont{Scagnoli}},
  \bibinfo{author}{\bibfnamefont{C.}~\bibnamefont{Mazzoli}},
  \bibinfo{author}{\bibfnamefont{E.}~\bibnamefont{Pomjakushina}},
  \bibinfo{author}{\bibfnamefont{K.}~\bibnamefont{Conder}}, \bibnamefont{and}
  \bibinfo{author}{\bibfnamefont{Y.}~\bibnamefont{Wang}},
  \bibinfo{journal}{Phys. Rev. Lett.} \textbf{\bibinfo{volume}{103}},
  \bibinfo{pages}{077602} (\bibinfo{year}{2009}).

\bibitem[{\citenamefont{Bland et~al.}()\citenamefont{Bland, de~Groot, Beale,
  Yakhou, Scagnolu, Dhesi, Angst, and Hatton}}]{bland}
\bibinfo{author}{\bibfnamefont{S.~R.} \bibnamefont{Bland}},
  \bibinfo{author}{\bibfnamefont{J.}~\bibnamefont{de~Groot}},
  \bibinfo{author}{\bibfnamefont{T.~A.~W.} \bibnamefont{Beale}},
  \bibinfo{author}{\bibfnamefont{F.}~\bibnamefont{Yakhou}},
  \bibinfo{author}{\bibfnamefont{V.}~\bibnamefont{Scagnolu}},
  \bibinfo{author}{\bibfnamefont{S.~S.} \bibnamefont{Dhesi}},
  \bibinfo{author}{\bibfnamefont{M.}~\bibnamefont{Angst}}, \bibnamefont{and}
  \bibinfo{author}{\bibfnamefont{P.~D.} \bibnamefont{Hatton}},
  \bibinfo{howpublished}{In draft}.

\bibitem[{\citenamefont{Blue et~al.}(2007)\citenamefont{Blue, Bates, Laing,
  Maneuski, O'Shea, Clarke, Prydderch, Turchetta, Arvanitis, and
  Bohndiek}}]{Blue:287}
\bibinfo{author}{\bibfnamefont{A.}~\bibnamefont{Blue}},
  \bibinfo{author}{\bibfnamefont{R.}~\bibnamefont{Bates}},
  \bibinfo{author}{\bibfnamefont{A.}~\bibnamefont{Laing}},
  \bibinfo{author}{\bibfnamefont{D.}~\bibnamefont{Maneuski}},
  \bibinfo{author}{\bibfnamefont{V.}~\bibnamefont{O'Shea}},
  \bibinfo{author}{\bibfnamefont{A.}~\bibnamefont{Clarke}},
  \bibinfo{author}{\bibfnamefont{M.}~\bibnamefont{Prydderch}},
  \bibinfo{author}{\bibfnamefont{R.}~\bibnamefont{Turchetta}},
  \bibinfo{author}{\bibfnamefont{C.}~\bibnamefont{Arvanitis}},
  \bibnamefont{and} \bibinfo{author}{\bibfnamefont{S.}~\bibnamefont{Bohndiek}},
  \bibinfo{journal}{Nuc. Inst. Meth. A} \textbf{\bibinfo{volume}{581}},
  \bibinfo{pages}{287} (\bibinfo{year}{2007}).

\bibitem[{\citenamefont{Blue et~al.}(2009)\citenamefont{Blue, Clarke, Houston,
  Laing, Maneuski, Prydderch, Turchetta, and O'Shea}}]{Blue:215}
\bibinfo{author}{\bibfnamefont{A.}~\bibnamefont{Blue}},
  \bibinfo{author}{\bibfnamefont{A.}~\bibnamefont{Clarke}},
  \bibinfo{author}{\bibfnamefont{S.}~\bibnamefont{Houston}},
  \bibinfo{author}{\bibfnamefont{A.}~\bibnamefont{Laing}},
  \bibinfo{author}{\bibfnamefont{D.}~\bibnamefont{Maneuski}},
  \bibinfo{author}{\bibfnamefont{M.}~\bibnamefont{Prydderch}},
  \bibinfo{author}{\bibfnamefont{R.}~\bibnamefont{Turchetta}},
  \bibnamefont{and} \bibinfo{author}{\bibfnamefont{V.}~\bibnamefont{O'Shea}},
  \bibinfo{journal}{Nucl. Instr. Meth. A} \textbf{\bibinfo{volume}{604}},
  \bibinfo{pages}{215} (\bibinfo{year}{2009}).

\bibitem[{MI3()}]{MI3}
\bibinfo{howpublished}{\url{http://mi3.shef.ac.uk}}.

\end{thebibliography}
\end{document}